\documentclass[11pt, a4paper]{article}
\usepackage{jcappub}
\usepackage{bm}

\newcommand{\Mpc}{{\;\rm Mpc}}

\newcommand{\keV}{{\;\rm keV}}

\title{The role of the \emph{e}ROSITA all-sky survey in searches for sterile
neutrino dark matter}
\author{Fabio Zandanel,}
\author{Christoph Weniger,}
\author{and Shin'ichiro Ando}
\emailAdd{f.zandanel@uva.nl}
\emailAdd{c.weniger@uva.nl}
\emailAdd{s.ando@uva.nl}
\affiliation{GRAPPA Institute, University of Amsterdam, 1098 XH
Amsterdam, The Netherlands}
\date{\today}

\abstract{We investigate for the first time the potential of angular auto- and cross-correlation
power spectra in identifying sterile neutrino dark matter in the cosmic X-ray
background.  We take as reference the performance of the soon-to-be-launched
\emph{e}ROSITA satellite.  The main astrophysical background sources against
sterile neutrino decays are active galactic nuclei, galaxies powered by X-ray
binaries, and clusters of galaxies.  While sterile neutrino decays are always
subdominant in the auto-correlation power spectra, they can be efficiently
enhanced when cross-correlating with tracers of the dark matter distribution
such as galaxies in the 2MASS catalogues.  We show that the planned four-years
\emph{e}ROSITA all-sky survey will provide a large enough photon statistics to
potentially yield very stringent constraints on the decay lifetime, enabling to
firmly test the recently claimed 3.56-keV X-ray line found towards several
clusters and galaxies and its decaying dark matter interpretation. However, we
also show that in order to fully exploit the potential of \emph{e}ROSITA for
dark matter searches, it is vital to overcome the shot-noise limitations
inherent to galaxy catalogues as tracers for the dark matter distribution.}

\begin{document}
\maketitle

\section{Introduction}

Revealing the non-gravitational nature of dark matter particles is one of the
most important goals of modern astrophysics and cosmology.  Even though weakly
interacting massive particles (WIMPs) are among the most popular and
well-motivated candidates~\cite{Jungman:1995df, Bertone:2004pz}, so far no
established signature thereof has been found at colliders, direct, or indirect
searches. This urges the investigation of alternative possibilities. Sterile
neutrinos with keV masses are well-known and well-motivated candidates for
dark matter~\cite{Dodelson:1993je, Shi:1998km, Boyarsky:2009ix,2012PDU.....1..136B}.
Observationally, keV-mass sterile neutrinos are interesting as they behave as
warm dark matter (WDM). This can potentially mitigate some of the small-scales
problems of the cold dark matter (CDM) scenario, e.g., the so called missing
satellite and too-big-to-fail problems, or alternatively provide indirect
evidence for sterile neutrino dark matter (see, e.g.,
\cite{2014MNRAS.439..300L}). 

Sterile neutrinos are spin-1/2 singlets under the Standard Model gauge group,
and can decay into photon-neutrino pairs, $\nu_s \rightarrow \gamma \nu$.  A
clear smoking gun signal for sterile neutrino dark matter would be the
detection of monochromatic photons at half of the dark matter mass.  The
corresponding decay rate is proportional to the mixing angle $\theta^2$ between
active and sterile neutrinos, and is given by \cite{Pal:1981rm,
Boyarsky:2009ix}
\begin{equation}
    \Gamma_{\nu_s} \simeq (7.2\times 10^{29}~ {\rm s})^{-1}
     \left(\frac{\sin^2 2\theta}{10^{-8}}\right)
    \left(\frac{m_{\nu_s}}{1 \keV}\right)^5\;,
\end{equation}
where $m_{\nu_s}$ is the sterile neutrino mass.

Monochromatic lines from sterile neutrinos have been searched for in various
different targets, such as nearby galaxies and galaxy clusters
\cite{PhysRevD.74.033009, Boyarsky11072008, Boyarsky11092010,
2012JCAP...03..018W, PhysRevD.89.025017,2015arXiv150605519F}.  
Recently, there have been claims of
the detection of an unidentified 3.56-keV line from a stacked sample of galaxy
clusters~\cite{Bulbul:2014sua}, and from Andromeda and the Perseus
cluster~\cite{Boyarsky:2014jta}, with a subsequent number of works on the
issue, some of which confirmed the claim
\cite{2014arXiv1408.2503B,2014arXiv1408.4388B,2014arXiv1409.4143B,2015arXiv150805186I} and some not
\cite{2014arXiv1408.1699J, 2014PhRvD..90j3506M, 2014arXiv1408.4115A,
2014arXiv1411.0050U, 2014arXiv1411.1759J, 2015JCAP...02..009C,
2015PASJ...67...23T}.  This is an ongoing debate that deserves further
investigation (see, e.g., \cite{2014arXiv1411.0311L}).

\medskip

One modern and very promising approach to indirect dark matter searches with
photons at all wavelengths is the cross-correlations between the
electromagnetic signal and tracers of the dark matter
distribution~\cite{2013ApJ...771L...5C, Ando:2013xwa}. Up to now, these
searches were mostly related to dark matter annihilation in the gamma-ray regime, 
with few studies also on decaying dark matter \citep{2013ApJ...771L...5C, 2014arXiv1411.4651C,
2014FrP.....2....6F}. The main reason being that the focus of indirect dark matter searches 
so far has been mainly on WIMPs and their gamma-ray annihilation and decay products. However,
as already mentioned, the lack of any strong evidence for WIMPs have recently 
broadened the interest of the community, particularly after the 3.56-keV line claim
and its sterile neutrino dark matter interpretation. 

Cross-correlation searches aim at the dark matter
signal coming from cosmological distances, and have the advantage of
exploiting simultaneously spatial and spectral information from the full sky.
As tracers for dark matter in the local Universe, galaxy
catalogues~\cite{Ando:2013xwa} and cosmic shear (weak gravitational
lensing)~\cite{2013ApJ...771L...5C, Shirasaki:2014noa} proved to be very
promising. Additionally, with a tomographic approach one can further exploit
the redshift information, which allows a better discrimination between various astrophysical and dark matter related photon
sources~\cite{Ando:2014aoa, 2014arXiv1411.4651C, 2015ApJS..217...15X}.
The power of such cross-correlation searches manifested recently in the first
discovery of cross-correlations of the \emph{Fermi}-Large Area Telescope gamma-ray sky with the
CMB~\cite{2015ApJ...802L...1F} and galaxy catalogues~\cite{Xia:2015wka}, with
important implications for WIMP dark matter searches studied in
Refs.~\cite{Regis:2015zka, 2015arXiv150601030C}.  A recent exhaustive review
about the gamma-ray emission from cosmological distances, and the potential of
cross-correlation studies, can be found in Ref.~\cite{2015arXiv150202866F} (see
also Ref.~\cite{Sefusatti:2014vha} for a detailed discussion about the expected
WIMP dark matter signal).

\medskip

In the present paper, we investigate for the first time the potential of an angular power 
spectrum analysis of the cosmic X-ray background (CXB) in identifying sterile neutrino
dark matter.  The methods that we present are very general and can be used for
all scenarios with decaying dark matter.  We will show that the
auto-correlation power spectrum of X-ray emission from sterile neutrinos is
completely dominated by background sources, in particular by active galactic
nuclei (AGNs) and clusters of galaxies. However, as we will demonstrate, the
\emph{cross-correlation} with tracers of the dark matter distribution in the
Universe, like large galaxy catalogues, can greatly enhance the sensitivity
towards a sterile neutrino contribution to the X-ray flux, and hence can be a
sensitive probe of dark matter signals.

For the purpose of such a CXB analysis, a deep and (nearly) all-sky survey with
good energy and angular resolution is a fundamental requirement. The last X-ray
all-sky survey has been performed by ROSAT (e.g., \cite{1999A&A...349..389V}).
Current X-ray instruments such as XMM-\emph{Newton} (e.g.,
\cite{2001A&A...365L..27T}) and \emph{Chandra} (e.g.,
\cite{2002PASP..114....1W}) have impressive performances and achievements.
However, they only operate in pointed observations. The next X-ray all-sky
survey will be performed by \emph{e}ROSITA~\cite{Merloni:2012}, whose launch is
scheduled for 2016, with a set of 6-month full-sky observations up to four
years.  Therefore, \emph{e}ROSITA will be the reference for our predictions.  

As an exemplary (and readily available) tracer for the nearby dark matter
distribution in the local Universe, we will adopt the 2MASS galaxy catalogue \cite{2MRS}.
It provides a nearly complete information on the distribution of
galaxies up to $z \approx 0.1$, with a sky coverage of 91\%.  We will discuss
also the potential ultimate reach of our method when other tracers are adopted.

\medskip

The paper is organised as follows. In section~\ref{sec:SigBack} we will discuss
in detail the different background and the signal contributions to the CXB.
Their auto-correlation power spectrum is discussed in
section~\ref{sec:autocorr}, whereas the cross-correlation power spectrum with
tracers of dark matter is discussed in section~\ref{sec:crosscorr}. In
section~\ref{sec:results} we present our projected sensitivity for future
instruments.  Eventually, in section~\ref{sec:conc}, we present our
conclusions.

If not specified otherwise, we adopt a halo mass $M_{\Delta}$ defined with
respect to a density that is $\Delta=200$ times the \emph{critical} density of
the Universe at redshift $z$. Throughout the paper, we assume a standard
$\Lambda$CDM cosmology with the parameters $\Omega_\Lambda=0.73$,
$\Omega_\mathrm{m}=0.27$, $\Omega_\mathrm{b}=0.05$, $\Omega_{\rm dm} = 0.22$,
and $H_0 = 100\, h \,\text{km}\;\text{s}^{-1}\,\text{Mpc}^{-1}$ with $h=0.7$.


\section{Signal and backgrounds in X-ray searches for sterile neutrinos} 
\label{sec:SigBack}
In the following three sections, we will discuss and characterise the mean
intensity of a sterile neutrino dark matter signal and of the most relevant
background components. These backgrounds are fluxes from AGNs, X-ray binaries
hosted in galaxies, and emission from clusters of galaxies.  

\subsection{Sterile neutrino dark matter}
The intensity of photons from sterile neutrino decay into $\gamma \nu$ final
states along the direction ${\bm n}$, defined as a number of photons received
per unit area, time, solid angle, and energy range, is given by\footnote{ The
lower limit for the redshift integration in Eqs.~(\ref{eqn:GammaFluxEG}) is
chosen to be $z = 0.003$.  The mean intensity, as well as the angular power
spectrum at high multipoles, are shown to be insensitive to this choice
(e.g.,~\cite{Ando:2013ff}). See also the window function dependence on $z$ in
the right panel of Fig.~\ref{fig:Cl_auto}.}
\begin{equation}
  \begin{split}
      I_{\nu_s}(E, \chi {\bm n}) =
    \frac{ \Gamma_{\nu_s}}{4\pi m_{\nu_s}}
    \int_0^\infty \frac{dz}{H(z)\,(1+z)^3}
    ~\rho_{\nu_s}(z, \chi {\bm n})
    ~\delta_{\rm D} \left[(1+z)E - \frac{m_{\nu_s}}{2}\right],
  \end{split}
  \label{eqn:GammaFluxEG}
\end{equation}
where $\delta_D$ is the Dirac delta function, $H(z)=H_0
\sqrt{\Omega_\mathrm{m}(1+z)^3 + \Omega_\Lambda}$ is the Hubble parameter as
function of redshift $z$, $\rho_{\nu_s}(z, \chi {\bm n})=\rho_{\rm dm}(z, \chi
{\bm n})$ denotes the mass density of sterile neutrino dark matter in the
direction $\bm n$ at a distance $z$, and $\chi$ is the co-moving distance.

We will first evaluate the mean intensity of extragalactic sterile neutrino
dark matter and hence the overall contribution to the CXB.  We note  that
adopting the ensemble average yields $\langle \rho_{\nu_s}(z, \chi {\bf n})
\rangle = (1+z)^3 \Omega_{\nu_s} \rho_c = (1+z)^3 \Omega_{\rm dm} \rho_c$,
where $\rho_c$ is the present critical density of the Universe. Therefore,
after taking the ensemble average of Eq.~(\ref{eqn:GammaFluxEG}), and
convolving with a normal distribution in energy in order to represent the
finite energy resolution of an X-ray telescope, the mean intensity due to the
sterile neutrino decay is given by
\begin{equation}
 I_{\nu_s}(E) = \int_0^\infty d\chi~
  W_{\nu_s}([1+z]E, z)\;,
  \label{eq:intensity nu_s}
\end{equation}
where the window function $W_{\nu_s}(E, z)$ is
\begin{equation}
 W_{\nu_s}(E, z) = \frac{\Omega_{\rm dm} \rho_c 
  \Gamma_{\nu_s}}{2(2\pi)^{3/2} m_{\nu_s} (1+z)\sigma_{E}}
  \exp\left[-\frac{(E - m_{\nu_s}/2)^2}{2(1+z)^2\sigma_{E}^2}\right]\;.
  \label{eq:window nu_s}
\end{equation}
We adopt $\sigma_{E} = 0.138 ~ {\rm keV} / \sqrt{8 \ln 2} = 0.0586 \keV$
corresponding to the energy resolution expected for \emph{e}ROSITA
\cite{Merloni:2012} at 6 keV.\footnote{\label{note1} Value taken from
http://www.mpe.mpg.de/455799/instrument.  We approximate the energy resolution
as being constant in our energy range of interest, while in reality it improves
at lower energies and worsen at higher energies. For example, in the case of
XMM-\emph{Newton} (see the XMM-\emph{Newton} Users' Handbook), the FWHM energy
resolution at 1, 6 and 10 keV is about 70, 150 and 180~eV, respectively.} As
expected, Eqs.~(\ref{eq:intensity nu_s}) and (\ref{eq:window nu_s}) show that,
for a given energy $E$, the dominant contribution come mainly from redshifts
around $z = m_{\nu_s}/2E - 1$.

\begin{figure}
    \centering
    \includegraphics[width=0.55\textwidth]{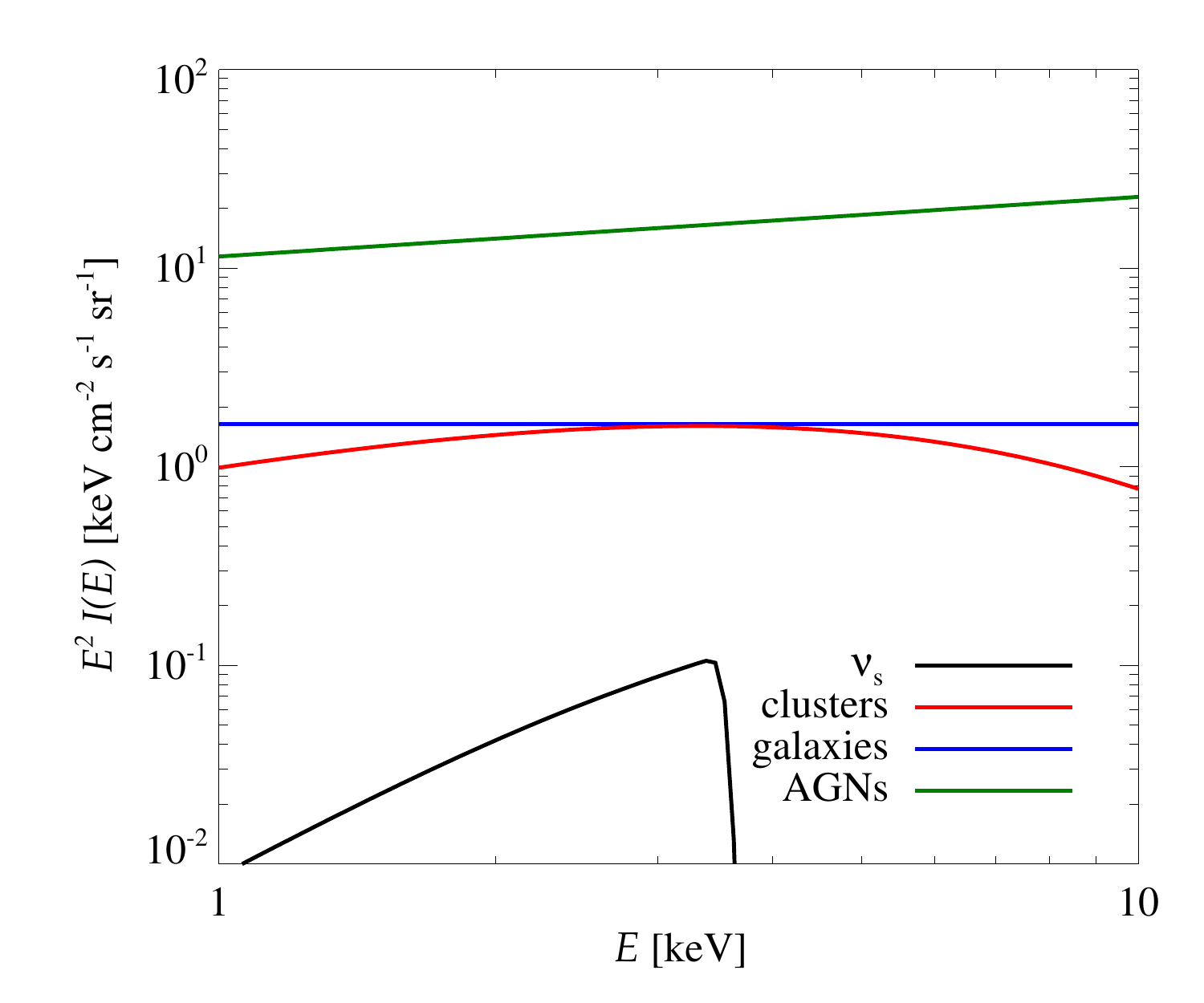}
    \caption{Contributions to the CXB mean intensity as a function of energy.
    We show the sterile neutrino signal for our reference scenario as described
    in the text (motivated by the findings of
    Refs.~\cite{Bulbul:2014sua,Boyarsky:2014jta}), compared with other
    astrophysical contributions: unresolved AGNs and galaxies, and galaxy
    clusters (both resolved and unresolved).}
    \label{fig:intensity}
\end{figure}

Fig.~\ref{fig:intensity} shows the mean intensity due to sterile neutrino
decays, for a mass $7.12$ keV and $\sin^22\theta= 7.6\times10^{-11}$, which
will be our \emph{reference scenario} motivated by the findings in
Refs.~\cite{Bulbul:2014sua,Boyarsky:2014jta}, compared with the CXB
contribution from AGNs, galaxies, and galaxy clusters (these background
components will be discussed in sections~\ref{sec:AGN}--\ref{sec:clusters}).
The sterile neutrino component is completely subdominant, being smaller by two
orders of magnitude with respect to the dominant AGN contribution, even at the
peak energy around 3.5~keV.

\subsection{Active galactic nuclei}
\label{sec:AGN}
AGNs are believed to provide the dominant contribution to the measured
CXB~\cite{1992ARA&A..30..429F, Treister:2006ec, Gilli:1999pf}.  We are here
interested in the contribute from \emph{unresolved} AGNs, since we can assume
that all resolved point sources can be masked before performing the angular
power-spectrum analysis.  The unresolved contribution is extrapolated from the
resolved one. Deep X-ray surveys suggest that the unresolved component
contributes 76\% to 82\% in the 0.5--2~keV and 2--8~keV energy bands to the
overall CXB, respectively~\cite{Lehmer:2012ak}.

We describe AGNs via their X-ray luminosity function (XLF) defined as the
co-moving number density per unit logarithmic luminosity range $\Phi_{\rm
AGN}(L_{\rm X}, z) \equiv dn_{\rm AGN} / d\log_{10} L_{\rm X}$. There are
several parameterisations for the XLF and we adopt the Luminosity And Density
Evolution (LADE) model~\cite{Aird2010}:
\begin{equation}
 \Phi_{\rm AGN}(L_{\rm X}, z) = K(z)
  \left[\left(\frac{L_{\rm X}}{L_\star(z)}\right)^{\gamma_1} +
   \left(\frac{L_{\rm X}}{L_\star(z)}\right)^{\gamma_2}\right]^{-1},
  \label{eq:XFL_AGN}
\end{equation}
where $K(z) = 10^{-4.53 - 0.19 (1+z)} \Mpc^{-3}$, $\gamma_1 = 0.62$, and
$\gamma_2 = 3.01$. The characteristic luminosity $L_\star(z)$, where the
power-law break happens, depends on redshift and is parameterised as
\begin{equation}
 L_\star(z) = L_0
  \left[\left(\frac{1+z_c}{1+z}\right)^{p_1} +
   \left(\frac{1+z_c}{1+z}\right)^{p_2}\right]^{-1},
  \label{eq:Lstar_AGN}
\end{equation}
where $L_0 = 10^{44.77}~\mathrm{erg~s^{-1}}$, $z_c = 0.75$, $p_1 = 6.36$, and
$p_2 = -0.24$. Note that this XLF refers to luminosities integrated between
$E_{\rm min} = 2\keV$ and $E_{\rm max} = 10\keV$ in the AGN rest frame.

The purpose of the current study is not to get a precise estimate of the mean
intensity of unresolved AGNs, but to show the power of angular fluctuation to
identify sterile neutrino dark matter.  Therefore, we simply assume that the
each AGN features a simple power-law spectrum, $E^{-\Gamma}$, with a spectral
index $\Gamma = 1.7$ in the relevant X-ray energy range (e.g.,
\cite{Tozzi:2006mj}).  With this assumption, the differential luminosity at an
energy $E$ (defined as a number of photon emitted per unit time and per unit
energy range), at redshift $z$, is
\begin{equation}
 \mathcal L_{\rm X} (E, z) = L_{\rm X} \frac{2-\Gamma}{E_{\rm
  max}^{2-\Gamma} -  E_{\rm max}^{2-\Gamma}} E^{-\Gamma}\;,
\end{equation}
where $L_X$ denotes the luminosity used in Eq.~\eqref{eq:XFL_AGN}.  The mean
AGN intensity is then calculated, similarly to Eq.~\eqref{eq:intensity nu_s},
as\footnote{The lower limit for the redshift integration in
Eq.~(\ref{eq:intensity_AGN}) for AGNs, as well as for galaxies in
section~\ref{sec:galaxies}, is chosen to be 10$^{-4}$. This does not impact the
calculations at all because all the AGNs and galaxies within this redshift give
fluxes above the eROSITA sensitivity and hence they are completely resolved.
See also the window function dependence on $z$ in the right panel of
Fig.~\ref{fig:Cl_auto}.}
\begin{equation}
 I_{\rm AGN}(E) = \int_0^\infty d\chi~W_{\rm AGN} ([1+z]
  E, z)\;,
  \label{eq:intensity_AGN}
\end{equation}
where the AGN window function is given as
\begin{equation}
 W_{\rm AGN}(E, z) = \frac{1}{4\pi\ln 10} \int_{L_{\rm X, min}}^{L_{\rm
  X, max}} \frac{dL_{\rm X}}{L_{\rm X}} \Phi_{\rm AGN}(L_{\rm X}, z)
  \mathcal L_{\rm X}(E, z)\;.
  \label{eq:window_AGN}
\end{equation}
For the lower and upper limit of luminosity integration, we adopt $L_{\rm X,
min} = 10^{41}~\mathrm{erg~s^{-1}}$ and
\begin{equation}
    L_{\rm X, max}(z) = \frac{4\pi d_L^2 F_{\rm sens}}{(1+z)^{2-\Gamma}}
    \frac{E_{\rm max}^{2-\Gamma} - E_{\rm min}^{2-\Gamma}}{\tilde E_{\rm
    max}^{2-\Gamma} - \tilde E_{\rm min}^{2-\Gamma}}\;,
\end{equation}
respectively (e.g.,~\cite{Aird2010}). $F_{\rm sens}$ is the flux sensitivity of
the considered X-ray telescope, and we adopt
$4.4\times10^{-14}$~erg~cm$^{-2}$~s$^{-1}$ for the energy range between $\tilde
E_{\rm min} = 0.5\keV$ and $\tilde E_{\rm max} = 2\keV$ for \emph{e}ROSITA
\cite{Merloni:2012}, which corresponds to the point-like sensitivity expected
after the first six months of survey.\footnote{This implies that we will obtain
slightly more conservative results than considering the sensitivity for the
full four year survey, but we tested that the change in our final constraints
is not significant (mostly since the result is limited by other factors,
see discussion below).} The mean intensity of unresolved AGNs computed
in this way is shown in Fig.~\ref{fig:intensity}. It is by far the dominant
contribution to the CXB. 

The AGN luminosity density of the adopted model has a $1\,\sigma$ uncertainty
below 50\% at low, $z\lesssim0.2$, and high, $z\gtrsim3$ redshifts, and below 
$20\%$ in the middle range~\cite{Aird2010}. Therefore, while, as explained above, 
we refrain from a precise estimation of the unresolved AGN contribution, we note that 
our conclusions are robust against the present uncertainty.

\subsection{Galaxies}
\label{sec:galaxies}
Galaxies are also X-ray sources as they host X-ray binaries.  The flux
distribution of galaxies approaches that of AGNs at the faint
end~\cite{Lehmer:2012ak}.  We adopt the XLF (in the rest-frame energy range
band 0.5--2~keV) from Ref.~\cite{2007ApJ...667..826P}, where the redshift
evolution was studied from a galaxy sample below $z \approx 1$.  The galaxy XLF
is well represented with the following log-normal distribution function:
\begin{equation}
 \Phi_{\rm gal}(L_{\rm X}, z) = \Phi_\star
 \left(\frac{L_{\rm X}}{L_\star(z)}\right)^{1-\alpha}
 \exp\left[-\frac{1}{2\sigma^2}\ln^2
      \left(1+\frac{L_{\rm X}}{L_\star(z)}\right)\right],
\end{equation}
where $\Phi_\star = 10^{-2.23}\Mpc^{-3}$, $\alpha = 1.43$, and $\sigma = 0.72$.
The characteristic luminosity $L_\star$ evolves with redshifts while
$\Phi_\star$ is constant (pure-luminosity evolution), and its evolution is
parameterised as $L_\star = 10^{39.74} ~\mathrm{erg~s^{-1}}
[(1+z)/1.25]^{1.9}$.

The CXB mean intensity $I_{\rm gal}(E)$, as well as its window function $W_{\rm
gal}(E, z)$, are computed in the same way as for the case of AGNs, by using the
galaxy XLF $\Phi_{\rm gal}(L_{\rm X}, z)$ in Eqs.~(\ref{eq:intensity_AGN}) and
(\ref{eq:window_AGN}).  This time, however, we adopt a $E^{-2}$
spectrum~\cite{Young:2012bm}, and thus the differential luminosity is given by
\begin{equation}
 \mathcal L_{\rm X}(E, z) = L_{\rm X}
  \frac{E^{-2}}{\ln(E_{\rm max}/E_{\rm min})},
\end{equation}
where $E_{\rm min} = 0.5\keV$ and $E_{\rm max} = 2\keV$, and the upper limit of
the luminosity integration is simply computed as $L_{\rm X, max}(z) = 4\pi
d_L^2 F_{\rm sens}$, with $F_{\rm sens} = 4.4 \times 10^{-14}
~\mathrm{erg~cm^{-2}~s^{-1}}$ as for AGNs.  The mean intensity of unresolved
galaxies computed in this way is shown in Fig.~\ref{fig:intensity}.  It is only
about 10\% of the emission from unresolved AGNs, but still an order of
magnitude above our reference sterile neutrino signal. Also in 
this case, the $1\,\sigma$ uncertainty on the luminosity density is 
below 50\% \cite{2007ApJ...667..826P} and, therefore,
our conclusions are robust.

\subsection{Clusters of galaxies}
\label{sec:clusters}
Galaxy clusters are prominent X-ray emitters as the ambient gas, the
intra-cluster medium (ICM), is a hot thermal plasma radiating bremsstrahlung
emission (see \cite{Voit:2005} for a review). Indeed, X-ray observations are
one of the main method to identify clusters. Clusters also represent some of
the most promising targets to identify sterile neutrino dark matter, or any
decaying dark matter candidate, as this process is directly proportional to the
halo mass and they indeed are the most massive halos in the Universe. 

Future X-ray instruments will help in narrowing down the contribute of galaxy
clusters to the CXB.  In fact, \emph{e}ROSITA is predicted to be able to
resolve \emph{all} galaxy clusters in the Universe \cite{Merloni:2012}.
However, differently from the previous cases of AGNs and galaxies, we must here
consider the whole population of resolved \emph{and} unresolved clusters and
their thermal X-ray emission. This is because one of the questions here is: do
clusters of galaxies also represent the bulk of the angular power spectrum for
the sterile neutrino dark matter, and, therefore, is their thermal
bremsstrahlung emission to be considered a dominant background in this
approach? We underline that the bremsstrahlung emission depends on the gas
density squared, which implies that significant morphological differences
between a dark matter decay signal and the gas emission are expected.

Following the above notation, the mean cluster X-ray intensity due to
bremsstrahlung radiation is given by\footnote{The lower limit for the redshift
integration in Eq.~(\ref{eqn:intensity_cl}) is chosen to be $z = 0.01$ as the
closest known galaxy clusters (see, e.g., \cite{Piffaretti2011}).}
\begin{equation}
  \begin{split}
      I_\mathrm{cl}(E) =
     \int_0^\infty d\chi
     ~W_\mathrm{cl}([1+z]E, z)
     \langle\rho_\mathrm{gas}^2\rangle\;,
  \end{split}
  \label{eqn:intensity_cl}
\end{equation}
where $\rho_\mathrm{gas}$ is the ICM density, and 
\begin{equation}
  \begin{split}
    \langle\rho_\mathrm{gas}^2\rangle = 
  \left( \frac{1}{\Omega_\mathrm{b}~\rho_\mathrm{c}} \right)^2 
  \int d M_{200} \frac{d n}{d M_{200}} 
  \int dV \rho_\mathrm{gas}^2(r|M_{200}) \;,
  \end{split}
  \label{eqn:gas2_cl}
\end{equation}
with integration boundaries that are discussed below.  The cluster
bremsstrahlung window function is
\begin{equation}
 W_\mathrm{cl}(E, z) = 
 \frac{\left( \Omega_{b}~\rho_{c} \right)^2}{4\pi}
 k_\mathrm{ff} ~\frac{ \left( k_\mathrm{B}T_\mathrm{gas} \right)^{-1/2} }{E} 
 ~\mathrm{exp} \left(- \frac{E}{k_\mathrm{B}T_\mathrm{gas}} \right) \;,
 \label{eq:window_cl}
\end{equation}
where $k_\mathrm{B}T_\mathrm{gas}$ is the ICM temperature in keV. We define
$k_\mathrm{ff} = 9.5\times 10^{32} g_\mathrm{ff}$ (adapted from
\cite{SarazinBook}), where $g_\mathrm{ff}=1.1$ is the Gaunt factor for the
free-free emission, such that $I_\mathrm{cl}(E)$ is expressed in
cm$^{-2}$~s$^{-1}$~sr$^{-1}$~keV$^{-1}$. 

Our expression for the bremsstrahlung emission do not consider atomic line transitions
in the ICM. This simplification is needed in order to treat the problem
analytically. Atomic line transitions depend on temperature and mass of the
cluster, and on the density squared as bremsstrahlung. The exclusion of atomic
lines will hence have no impact on our study of morphological differences in
the angular power spectrum for the different components. One might worry that
our side-band analysis of section~\ref{sec:results}, which takes into account
spectral information, is significantly affected by this approximation. However,
as we will see, even neglecting entirely galaxy clusters in the analysis does
not change much the final results. Hence, we believe that avoiding the
transition lines does not impact much on this first study that we attempt here.

The adopted halo mass function $dn/dM_{200}$ in Eq.~(\ref{eqn:gas2_cl}), which
will be later also used for the sterile neutrino angular power spectrum
calculation, is based on the Tinker formalism \cite{Tinker:2008} and obtained
through the online application from \cite{Murray:2013}.  In
Eq.~(\ref{eqn:gas2_cl}), the mass function is integrated above
$M_{200}^\mathrm{cl,min} \times h = 10^{14}$~$M_{\odot}$, while the volume
integral is within $R_{200}$. The transition between galaxy groups and galaxy
clusters is not sharp or well-defined, and with the chosen
$M_{200}^\mathrm{cl,min}$ we already include large galaxy groups in the
integration.

Finally, we need a prescription for the ICM density and temperature. We adopt
the phenomenological model of \cite{Zandanel:2014gas}, which is based on X-rays
observations, and allows to assign a gas density (and temperature) to any
galaxy cluster using its mass only, in such a way that the observed X-ray and
Sunyaev-Zel'dovich scaling relations are correctly reproduced within about 20\%. 
Following \cite{Zandanel:2014gas}, we also statistically divide the cluster population in
half merging clusters and half relaxed clusters, which are characterised by
centrally cored and centrally peaked gas densities, respectively. This
prescription needs however the clusters' mass as $M_{500}$ which we obtain from
$M_{200}$ using the method in \cite{Hu:2003}.  As a result, we find that the
mean intensity of galaxy clusters, as shown in Fig.~\ref{fig:intensity}, is at
the level of unresolved galaxies.

\section{Auto-correlations of the angular power spectrum}
\label{sec:autocorr}
Discussing the auto-correlation angular power spectrum of the different signal
and background contributions to the CXB is useful for identifying the
components that will yield the strongest contrast at a given angular scale
$\theta = \pi/\ell$, where $\ell$ is the multipole.  As usual, we define the
intensity angular power spectrum of the signal and background fluxes as
\begin{equation}
    C_\ell \equiv \langle |a_{\ell m}|^2 \rangle\;,
\end{equation}
where the $a_{\ell m}$ are given by the decomposition of the flux into
spherical harmonics $Y_{\ell m}$, namely
\begin{equation}
  a_{\ell m}=
  \int d\Omega_{\bm{n}}\, I(\bm{n})\, Y_{\ell m}^\ast (\bm{n})\;.
  \label{eqn:alm1}
\end{equation}
The auto-correlation angular power spectrum of a given source population $A$
($= \nu_s$, ${\rm AGN}$, ${\rm gal}$, $\mathrm{cl}$) is then computed
as~\cite{Ando:2006cr, Ando:2014aoa}
\begin{equation}
  C_\ell^A(E) = \int_0^\infty \frac{d\chi}{\chi^2}
  ~W_A([1+z]E, z)^2~P_A\left( k=\frac{\ell}{\chi},z \right),
  \label{eqn:ClForm}
\end{equation}
where $P_A(k, z)$ is the power spectrum of the source $A$ at wave number $k$
and redshift $z$.  In the halo model, the latter can be divided into one- and
two-halo terms, $P_{\rm{A}} = P_{\rm{A}}^{\rm{1h}} + P_{\rm{A}}^{\rm{2h}}$
(higher halo terms are negligible). In the case of dark matter decay, it is the
nonlinear matter power spectrum ($P_{\nu_s} = P_\delta$) modelled
as~\cite{Cooray:2002dia}
\begin{eqnarray}
    P_{\delta}^{\rm{1h}} & = & \left( \frac{1}{\Omega_{\rm dm}\rho_{\rm c}}
    \right)^2 \int dM_{200} \frac{dn}{dM_{200}} \left[ \int 4\pi r^2 dr
        \rho_{\rm dm}(r) \frac{\sin(kr)}{kr}  \right]^2 \, , \\ \nonumber
    P_{\delta}^{\rm{2h}} & = & \left[ \left( \frac{1}{\Omega_{\rm dm}\rho_{\rm
            c}} \right) \int dM_{200} \frac{dn}{dM_{200}} \, b(M_{200},z ) \int
        4\pi r^2 dr \rho_{\rm dm}(r) \frac{\sin(kr)}{kr} \right]^2 \,
    P_{\rm{lin}} (k,\chi) \, , 
\label{eqn:ps1h2h}
\end{eqnarray}
where the mass integration starts from $M_{200}^{\nu_s,{\rm
lim}} \times h =10^{6}$~M$_\odot$, the radial integration goes up to $R_{200}$,
$P_{\mathrm{lin}}(k,\chi)$ is the linear matter power spectrum (obtained 
from \cite{Murray:2013}), and $b(M_{200},z)$ is the linear bias
\cite{Tinker:2010my}.  The lower mass limit for the integration of the sterile
neutrinos, which represent effectively a WDM candidate, can be chosen to be
around $10^6$--$10^8$~M$_\odot$ (e.g.,~\cite{Viel:2013fqw,2015arXiv150701998B}).  
Since the precise choice of the lower mass limit does not affect the
auto- and cross-power spectra that we will discuss, we adopt for definiteness 
$M_{200}^{\nu_s,{\rm lim}} \times h= 10^6$~M$_\odot$. Note also that our $P_{\mathrm{lin}}(k,\chi)$ 
is obtained for a CDM scenario, but as the WDM halo mass function starts to drop below its 
CDM counterpart around $10^9$~M$_\odot$ \cite{2015arXiv150701998B}, this choice of 
$P_{\mathrm{lin}}$ has no impact on our conclusions. The negligible relevance of the
two above choices, $M_{200}^{\nu_s,{\rm lim}}$ and $P_{\mathrm{lin}}(k,\chi)$, 
is made clear by Fig.~\ref{fig:masses_contrib} which shows that
the dominant contribution to the sterile neutrino power spectrum comes from 
halos with $M_{200} \times h \gtrsim 10^{11}$~M$_\odot$.

In order to calculate the angular power spectrum of Eqs.~(\ref{eqn:ps1h2h}), we
need the distribution of dark matter in halos $\rho_{\rm dm}$. For the present
study, we will adopt the Navarro-Frenk-White (NFW) profile~\cite{NFW:1997},
which is given by
\begin{equation}
    \rho_{\rm dm}(r) =
    \frac{\rho_s}{\left( r/r_s \right) \left( r/r_s +1 \right)^2} \; ,
\end{equation}
where $\rho_s$ and $r_s$ denote the scale density and radius, respectively.  We
use throughout the parameters $M_{200}$ and $R_{200}$ to indicate the halo mass
and radius, respectively.  Hence, our scale radius is defined as $r_s =
R_{200}/c_{200}$, where $c_{200}(M_{200},z)$ is the concentration parameter
\cite{Prada:2012}.  With this, one can show that, for a given halo mass
$M_{200}$, the scale density is given by
\begin{equation}
    \rho_s = \frac{M_{200}}{4 \pi r_s^3} \left[ \mathrm{ln}(1+c_{200}) -
      \frac{c_{200}}{1+c_{200}} \right]^{-1} \; .
\end{equation} 

In the case of clusters of galaxies, $P_\mathrm{Cl}(k,\chi)$ is similar to that
of Eqs.~(\ref{eqn:ps1h2h}), but with $M_{200}^{\rm
cl,lim} \times h =10^{14}$~M$_\odot$.  Additionally, $\rho_{\rm dm}(r)$ has to
be substituted by $\rho_\mathrm{gas}(r)^2$ and $(1/\Omega_{\rm dm}\rho_{\rm
c})$ by $(1/\Omega_{\rm b}\rho_{\rm c})^{2}$, and we assume that $P_{\rm
lin}(k,\chi)$ is related to the cluster power spectrum via the linear bias.

\begin{figure}
    \begin{center}
        \includegraphics[width=0.45\textwidth]{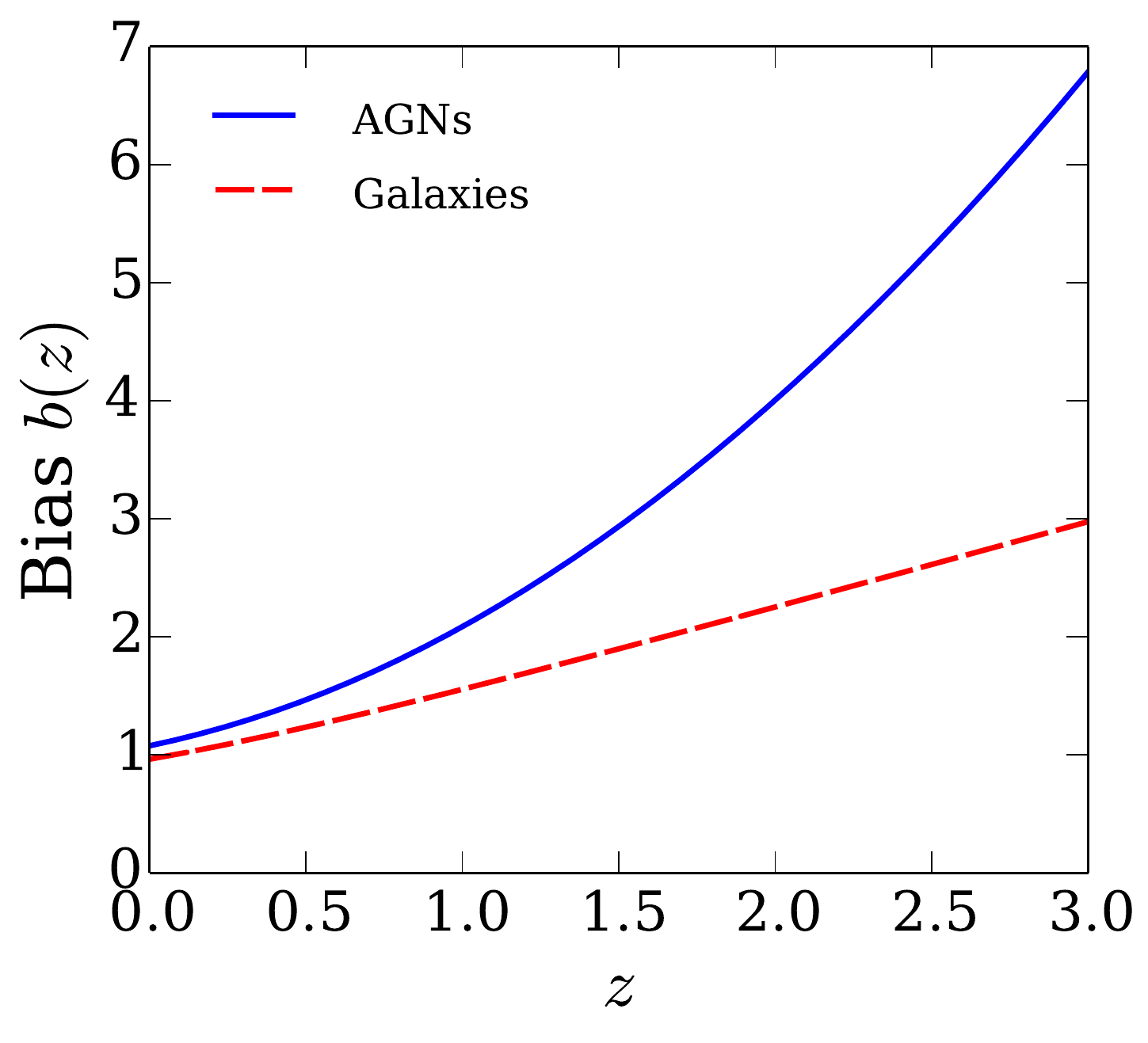}
        \caption{Bias of the X-ray AGNs and galaxies as a
        function of redshift $z$.}
        \label{fig:bias}
    \end{center}
\end{figure}

We assume that AGNs and galaxies are good tracers of the dark matter density,
and thus $P_\mathrm{AGN,gal}(k, z) = b_\mathrm{AGN,gal}^2(z) P_\delta(k, z)$,
where we again use Eq.~(\ref{eqn:ps1h2h}) for the nonlinear power spectrum of
dark matter density fluctuation $\delta$. For the AGNs, we adopted the halo
linear bias from \cite{Tinker:2010my} and assume that the X-ray AGNs reside in
dark matter halos with mass of $10^{13.1} M_\odot$~\cite{Allevato:2011jp}.  For
the X-ray emitting galaxies, we adopt the prescription given in \cite{Cappelluti:2012rd}. 
The bias models for both AGNs and galaxies are shown in Fig.~\ref{fig:bias} as a function of redshift. 
The uncertainty on the AGN bias can be estimated to be below about 10\%, while for
the halo mass in which X-ray AGNs are supposed to reside it is of about 30\%~\cite{Allevato:2011jp}.
The model for the bias of the X-ray galaxies of Ref.~\cite{Cappelluti:2012rd} well
reproduces measurements of star-forming and high-z Ly-break galaxies which
in turn have uncertainties below about 20\%. Note that the galaxy bias is
very close to $1$ for $z\lesssim1$, which is the redshift range most interesting
for dark matter searches.

Finally, since AGNs and galaxies are point-like sources, there is an additional
shot-noise contribution to the angular power spectrum, which is independent of
angular scale $\ell$. This {\it Poisson} term is computed by
(e.g.,~\cite{Ando:2006cr})
\begin{equation}
 C_{\rm P}^{\rm AGN, gal}(E) = \frac{1}{(4\pi)^2 \ln 10}\int_0^\infty
  \frac{d\chi}{\chi^2} 
  \int_{L_{\rm X, min}}^{L_{\rm X, max}} \frac{dL_{\rm X}}{L_{\rm X}}
  \Phi_{\rm AGN, gal}(L_{\rm X}, z) \mathcal L_{\rm X}(E, z)^2.
  \label{eq:Poisson}
\end{equation}

So far, we discussed only the {\it differential} (in energy) angular power
spectrum, but the same argument applies to the angular power spectrum {\it
integrated} over a certain energy range.  One first integrates the window
function $W_A$ over the relevant energies, and then computes the power spectrum
using Eq.~(\ref{eqn:ClForm}).  Similarly, for the Poisson term, one integrates
$\mathcal L_{\rm X}$ over the chosen energy range and then computes the shot
noise with Eq.~(\ref{eq:Poisson}). In order to emphasise the difference, we
will in the following explicitly show the energy dependence, $C_\ell(E)$, when
refereeing to differential power spectra, while the energy dependence is
suppressed, $C_\ell$, for the integrated quantities.

\begin{figure}
  \begin{center}
    \includegraphics[width=0.49\textwidth]{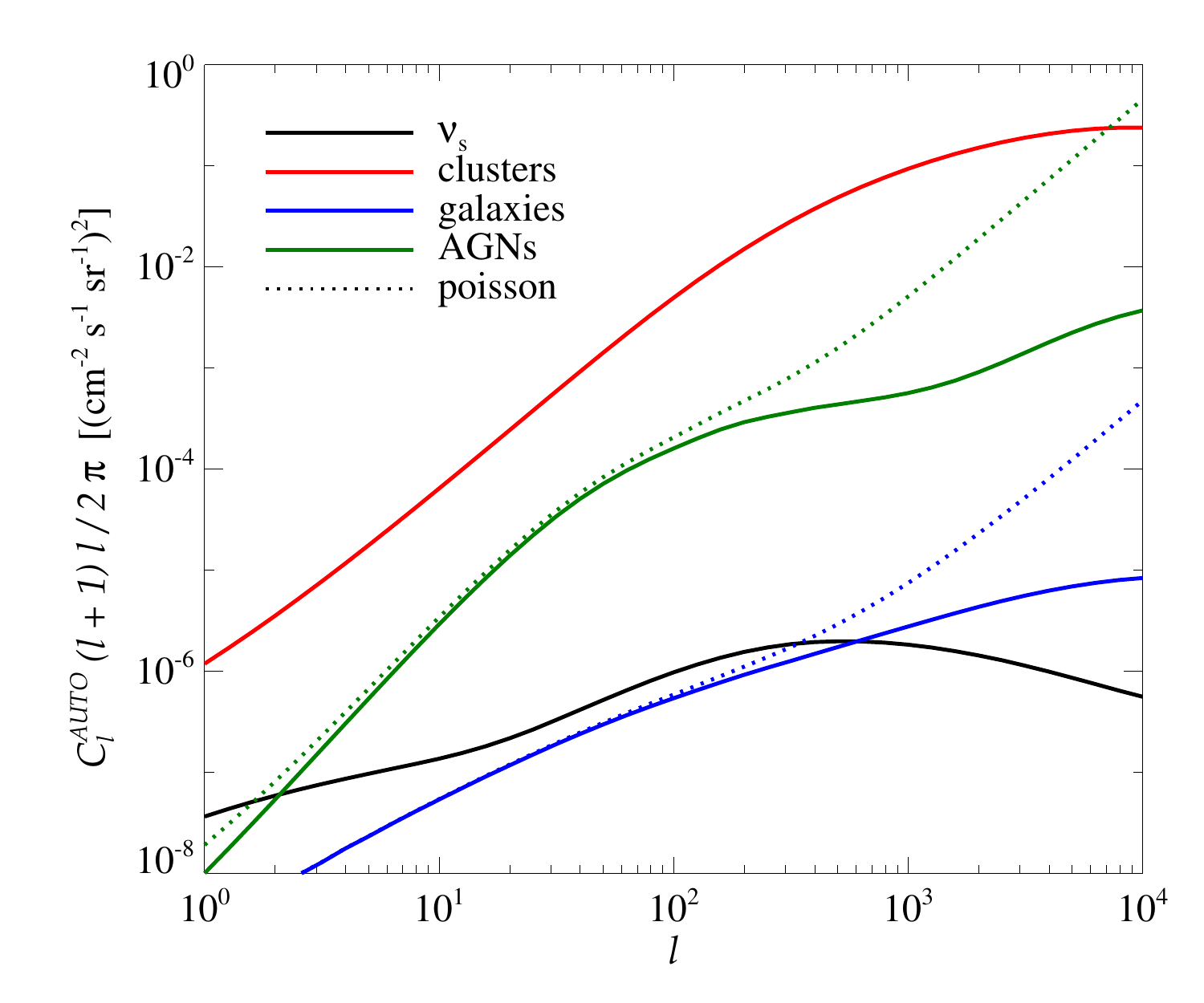}
    \includegraphics[width=0.49\textwidth]{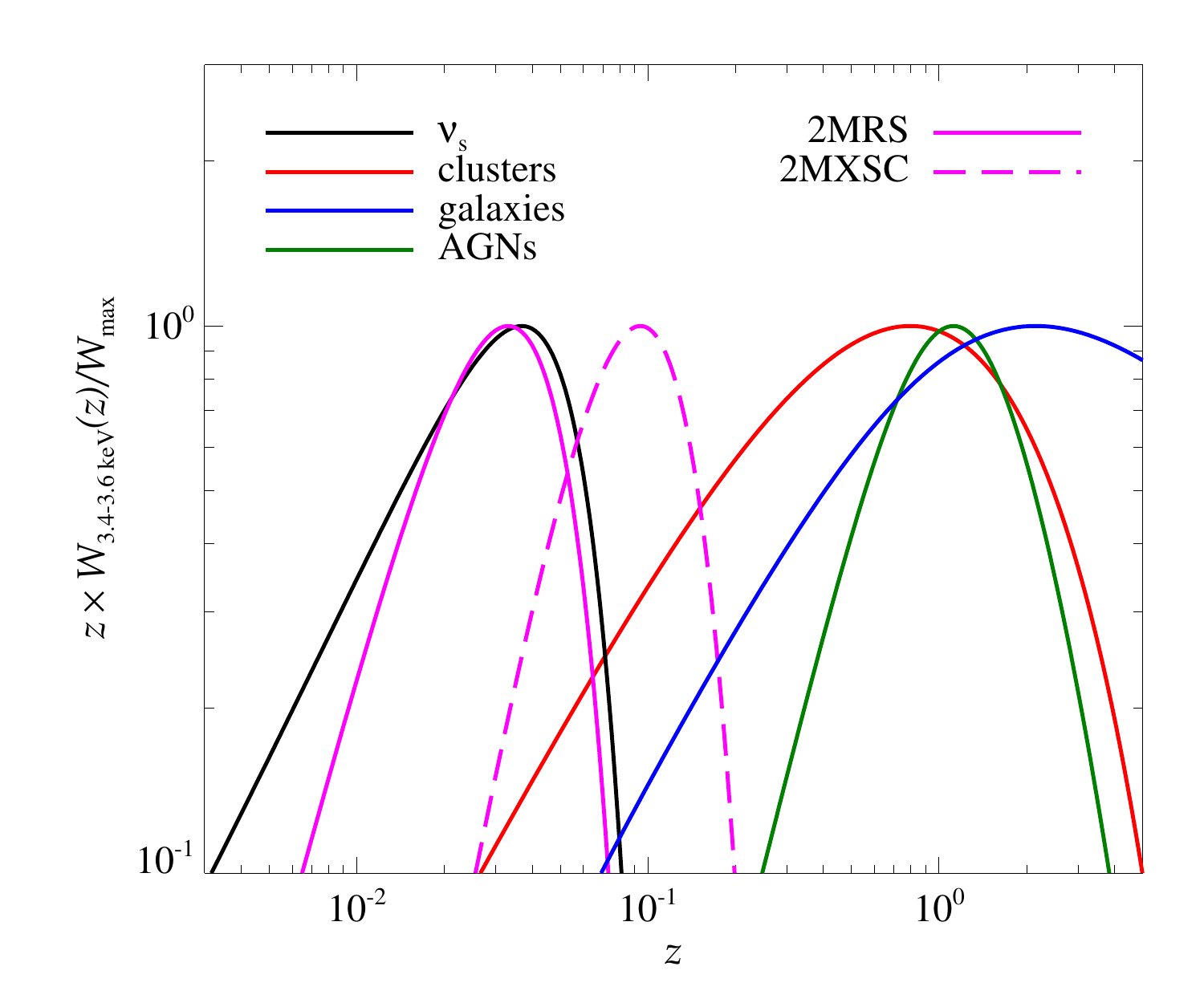}
    \caption{\textbf{Left.} Auto-correlation angular power spectrum of CXB due
    to sterile neutrino decays for our reference model, (unresolved) AGNs,
    (unresolved) galaxies, and (resolved and unresolved) clusters of galaxies
    in the $3.4-3.6$~keV energy band. For AGNs and galaxies, both the
    correlation term (solid) and total including Poisson term (dotted) are
    shown.  \textbf{Right.} Redshift-weighted window functions integrated in
    the $3.4-3.6$~keV energy band and normalised at their maximum for sterile
    neutrino decays with our reference model, AGNs, galaxies, 
    clusters of galaxies (fixing the ICM temperature to 5~keV), 
    and 2MRS and 2MXSC catalogues.}
    \label{fig:Cl_auto}
  \end{center}
\end{figure}

In the left panel of Fig.~\ref{fig:Cl_auto}, we show the auto-correlation
angular power spectrum integrated over the 3.4--3.6~keV energy band, adopting
our reference model for the sterile neutrino decay. The component coming from
sterile neutrinos is completely subdominant with respect to contributions from
clusters at all multipoles, as well as from AGNs, and, to a lesser extent, from
galaxies, for most angular ranges. Furthermore, we also show that at larger
multipoles the contributions from the Poisson term become increasingly
important and start to dominate. In the right panel of Fig.~\ref{fig:Cl_auto},
we show the shape of the window function integrated in the 3.4--3.6~keV energy
band for the different components, and for the 2MRS and 2MXSC galaxy catalogues
that will be introduced in the next section.

\begin{figure}
    \begin{center}
        \includegraphics[width=0.5\textwidth]{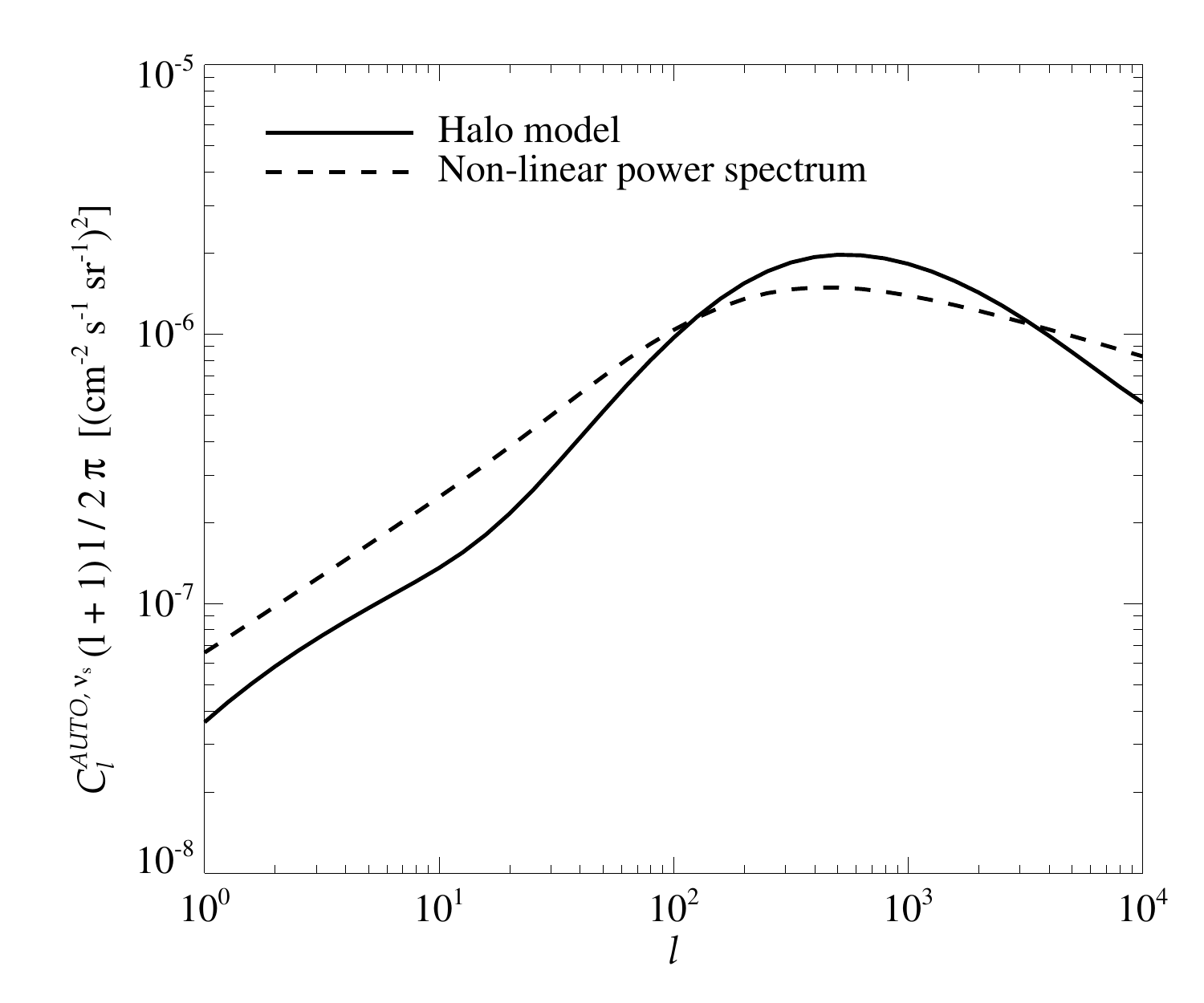}
        \caption{The auto-correlation angular power spectrum of sterile
        neutrino decays computed with the halo model (used in our main
        analysis) and with the non-linear matter power spectrum from CAMB,
        respectively.}
        \label{fig:HaloVScamb}
    \end{center}
\end{figure}

Actually, the power spectrum of the sterile neutrino component can be obtained
by using a non-linear matter power spectrum directly calculated with, e.g., the
Code for Anisotropies in the Microwave Background (CAMB; http://camb.info/).
We adopted the halo model approach in order to have control on the spatial part
of the angular power spectrum calculation for clusters of galaxies, and kept
this approach in all cases for consistency. In Fig.~\ref{fig:HaloVScamb}, we
compare the sterile neutrino auto-correlation angular power spectrum obtained
with the halo model approach to that obtained with the non-linear matter power spectrum from
CAMB and find reasonably good agreement, to within a factor of less than two.

\section{Cross correlations with the 2MASS galaxy catalogues}
\label{sec:crosscorr}
We will now show that, since the signal from sterile neutrino decays follows
the distribution of dark matter in the Universe, a cross-correlation between
the X-ray signal and tracers of the dark matter distribution is a very
promising strategy to isolate the dark matter contribution. To this end, we
will compute the cross-correlation with large galaxy catalogues from the 2MASS
survey~\cite{2MRS}, which in fact trace the dark matter distribution in the
\emph{local} Universe.

The 2MASS catalogue provides nearly complete information on the galaxy
distribution for the local Universe. The 2MASS Redshift Survey
(2MRS)~\cite{2MRS} is based on spectroscopic redshift determination of $N_{\rm
2MRS} \approx 43500$ galaxies up to $z \approx 0.1$ with a sky coverage of
91\%. The 2MRS galaxies follow approximately the redshift distribution given by
\begin{equation}
 \frac{dN_{\rm 2MRS}}{dz} \propto z
  \exp\left(-\left[\frac{z}{z_{\rm 2MRS}}\right]^2\right)\,,
\end{equation}
where $z_{\rm 2MRS} = 0.033$.  For comparison, we also adopt the 2MASS Extended
Source Catalogue (2MXSC), an even larger sample that contains $N_{\rm 2MXSC}
\approx 770000$ galaxies with a sky coverage of 67\%~\cite{Jarrett:2000qt,
Xia:2011ax}, but with less accurate redshift determination through photometry.
The redshift distribution of 2MXSC is approximately given by~\cite{Xia:2011ax}
\begin{equation}
 \frac{dN_{\rm 2MXSC}}{dz} \propto z^{1.9}
  \exp\left(-\left[\frac{z}{z_{\rm 2MXSC}}\right]^{1.75}\right)\,,
\end{equation}
where $z_{\rm 2MXSC} = 0.07$. We fix the normalisation of both the 2MRS and
2MXSC catalogues such that the redshift integration gives the total number
galaxies $N_{\rm 2MRS}$ and $N_{\rm 2MXSC}$, respectively. 

The angular cross-power spectrum of a given source population $A$ ($= \nu_s$,
${\rm AGN}$, ${\rm gal}$, $\mathrm{cl}$) with the 2MRS or 2MXSC galaxies,
labelled B, is therefore computed as~\cite{Ando:2006cr}
\begin{equation}
    C_\ell^{\rm A, B}(E) = \int
    \frac{d\chi}{\chi^2}W_{\rm A}([1+z]E, z)
    W_{\rm B}(\chi)P_{\rm A,B}
    \left(k = \frac{\ell}{\chi}, \chi\right),
    \label{eq:angular cross-power nu_s}
\end{equation}
where $W_{\rm B}(\chi)= (dz/d\chi) (dN_{\rm B}/N_{\rm B}dz)$ is the galaxy
catalogue window function (normalised to unity), and $P_{\rm A,B}(k, z)$ is the
cross-power spectrum.  For the cross-power spectrum, we adopt for the galaxy
catalogues bias $b_{\rm 2MRS/2MXSC} = b_{\rm gal}$.  In case of
cross-correlation with clusters of galaxies, the cross-power spectrum, $P_{\rm
cl,B} = P_{\rm cl,B}^{\rm 1h} + P_{\rm cl,B}^{\rm 2h}$, reads
\begin{eqnarray}
  \label{eqn:corss_ps1h}
  P_{\rm cl, B}^{\rm{1h}} & = & \left( \frac{1}{\Omega_{\rm b}\rho_{\rm c}}
  \right)^2 \frac{b_{\rm gal}(z)}{\Omega_{\rm dm}\rho_{\rm c}} \int dM_{200}
  \frac{dn}{dM_{200}} \\ 
  &  &\times \left[ \int 4\pi r^2 dr \rho_{\rm gas}^2(r) \frac{\sin(kr)}{kr}
  \right]\left[ \int 4\pi r^2 dr \rho_{\rm dm}(r) \frac{\sin(kr)}{kr}  \right]
  \nonumber \, ,\\
  \label{eqn:corss_ps2h}
  P_{\rm cl, B}^{\rm{2h}} & = & \left[ \left( \frac{1}{\Omega_{\rm b}\rho_{\rm
      c}} \right)^2 \int dM_{200} \frac{dn}{dM_{200}} \, b(M_{200},z ) \int
    4\pi r^2 dr \rho_{\rm gas}^2(r) \frac{\sin(kr)}{kr}\right]  \nonumber \\
  & &  \times \left[ \frac{b_{\rm gal}(z)}{\Omega_{\rm dm}\rho_{\rm
    c}} \int dM_{200} \frac{dn}{dM_{200}} \, b(M_{200},z )
    \int 4\pi r^2 dr \rho_{\rm dm}(r) \frac{\sin(kr)}{kr}
  \right]\, P_{\rm{lin}} (k,\chi) \, . 
\end{eqnarray}
In the case of cross correlation with sterile neutrino dark matter, the
cross-power spectrum reads as in Eqs.~(\ref{eqn:corss_ps1h}) and
(\ref{eqn:corss_ps2h}), but one has to substitute $\rho_{\rm gas}^2(r)$ with
$\rho_{\rm dm}(r)$ and $(1/\Omega_{\rm b}\rho_{\rm c})^{2}$ with
$(1/\Omega_{\rm dm}\rho_{\rm c})$.  For the cross-correlation with X-ray AGNs
or galaxies, we have $P_{\rm AGN/gal,B}(k, z) = b_{\rm AGN/gal}(z) b_{\rm
2MRS/2MXSC}(z) P_\delta (k, z)$.

\begin{figure}
  \centering
  \includegraphics[width=0.49\textwidth]{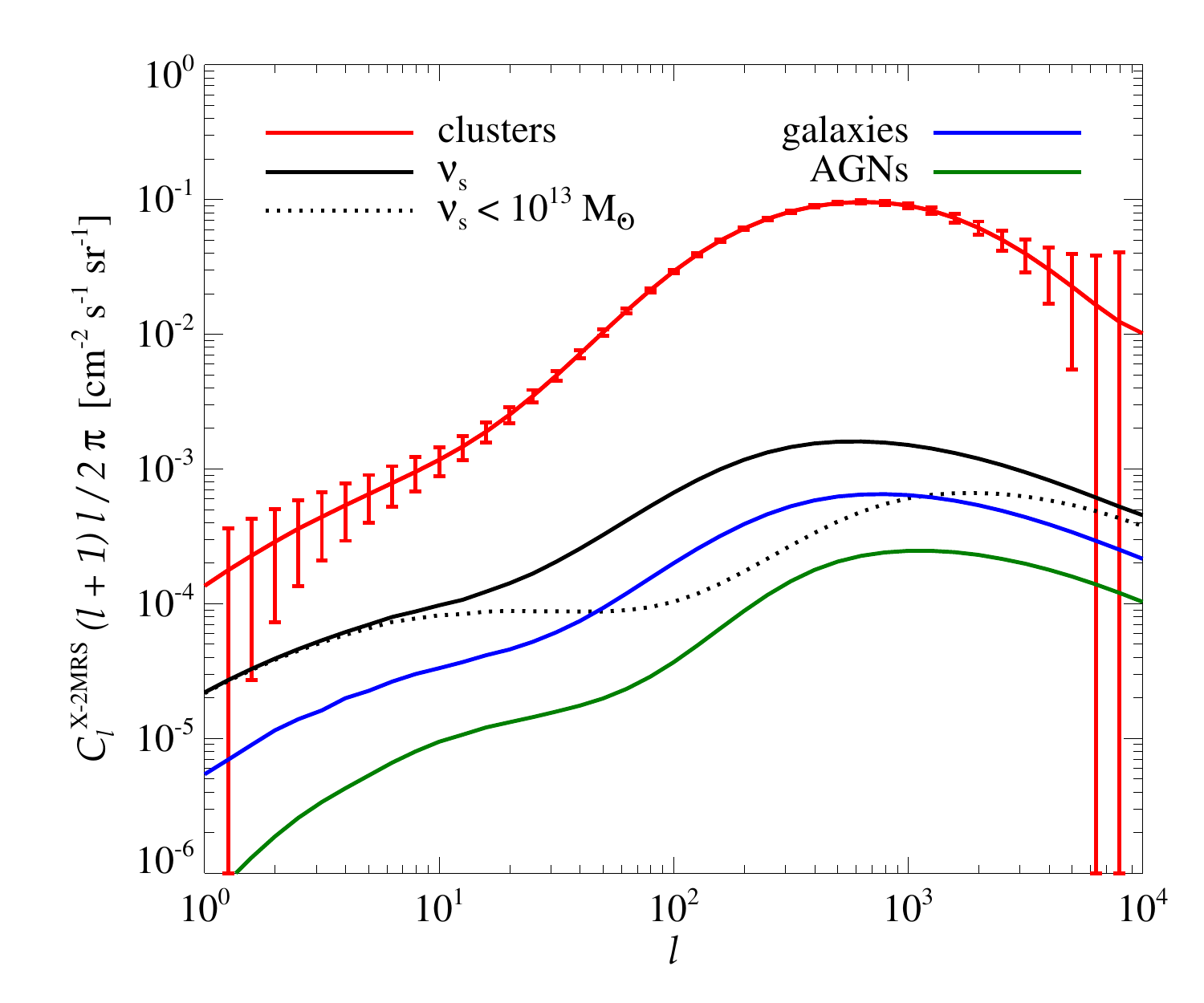}
  \includegraphics[width=0.49\textwidth]{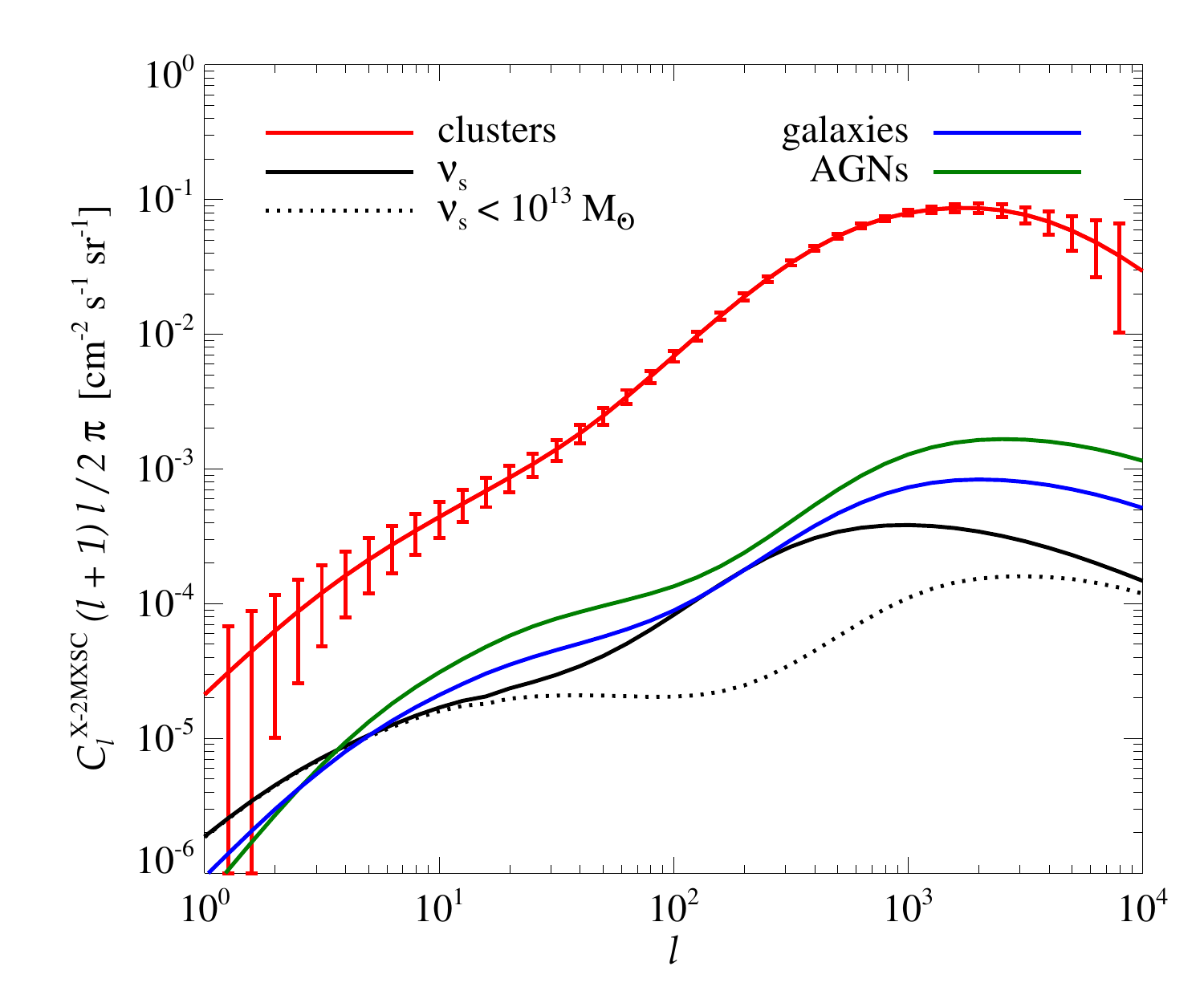}
  \caption{Cross-correlation angular power spectrum of CXB due to sterile
  neutrino decays for the reference model, (unresolved) AGNs, (unresolved)
  galaxies, and (resolved and unresolved) clusters of galaxies in the
  3.4--3.6~keV energy band. For sterile neutrinos, we also show the case when the
  integration upper mass limit is fixed to $M_{200}^{\nu, \mathrm{lim}} \times h = 10^{13}$~M$_\odot$ 
  (see main text for details). We overlay to the cluster component the 
  error bars $\Delta\,C_\ell/\sqrt{\Delta\,\ell}$, with $\Delta\,\ell=\ell/2$ the bin size, 
  estimated from the diagonal part of the covariance matrix for the total 
  power of Eq.~\ref{eq:error2}.  {\bf Left:} Cross correlation with 2MRS.
  {\bf Right:} Cross correlation with 2MXSC.}
  \label{fig:Cl_cross_AGN_gal}
\end{figure}

We show in Fig.~\ref{fig:Cl_cross_AGN_gal} the cross-correlation angular power
spectrum, $C_\ell^{\rm A,B}$, cross-correlated with 2MRS (left panel) and 2MXSC
(right panel) galaxies, integrated over 3.4--3.6~keV, and compare the sterile
neutrino component with AGNs, galaxies and clusters. The shapes of the power
spectra are similar for all components, as they trace the dark matter
distribution quite well, while some deviations are observed for clusters where
the emission depends on the gas density squared.

We find that the cross-correlation signal of (resolved and unresolved) galaxy
clusters dominates over all components at all multipole scales. At the same
time, the sterile neutrino component, subdominant in the auto-correlation, is
now comparable to the AGN and galaxy components. In particular, the sterile
neutrino component dominates over the AGNs and galaxies over all multipoles for
the cross correlation with 2MRS.  When cross-correlating with 2MXSC, AGNs and
galaxies dominate over the sterile neutrino component at all multipoles above
$l \approx 5$, while being still comparable. This can be explained by the fact
that unresolved AGNs and galaxies grow in number with redshift, therefore
correlating more with the higher redshift sample of 2MXSC. At the same time,
most of the power from sterile neutrinos comes from low redshifts, which causes
the 2MXSC cross-correlation power spectrum to be lower with respect to the 2MRS
case.  This can be seen also looking at the redshift variations of the window
functions of different components in the right panel of Fig.~\ref{fig:Cl_auto}.

\begin{figure}[t!]
  \centering
  \includegraphics[width=0.49\textwidth]{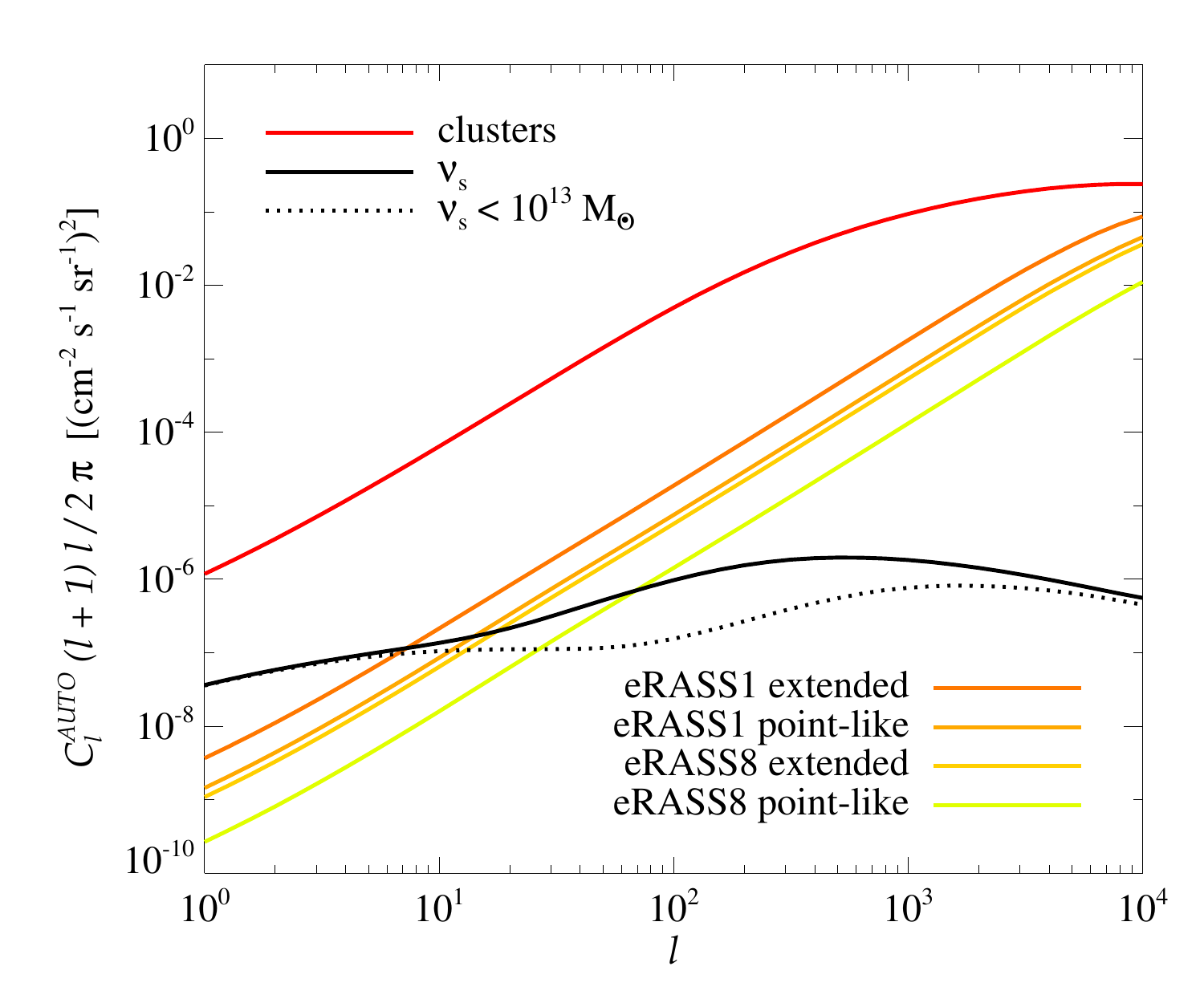}
  \includegraphics[width=0.49\textwidth]{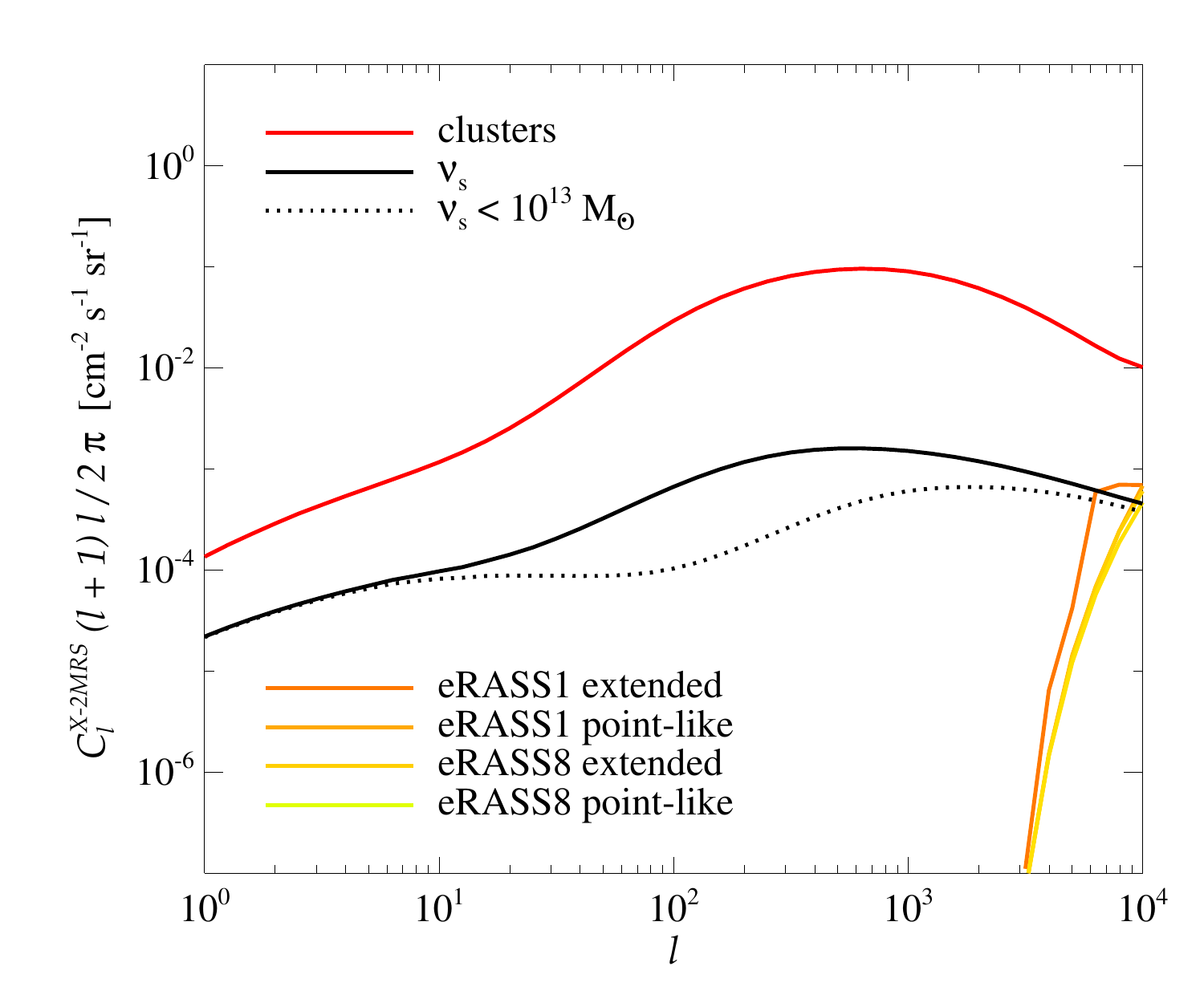}
  \caption{{\bf Left:} Auto-correlation power spectrum of sterile neutrino
  decays, for the reference model, and clusters of galaxies in the 3.4--3.6~keV
  energy band. For clusters of galaxies, we show the case when all resolved and
  unresolved objects are included, and the cases where we exclude the objects
  exceeding the \emph{e}ROSITA expected sensitivities (for point-like/extended
  sources and eRASS1/eRASS8 results). For sterile neutrinos, we also show the
  case when the integration upper mass limit is fixed to
  $M_{200}^{\nu, \mathrm{lim}} \times h = 10^{13}$~M$_\odot$. {\bf Right:} The same but for cross-correlation
  with 2MRS.}
  \label{fig:Cl_eRASSlimits}
\end{figure}

\begin{figure}[t]
  \centering
  \includegraphics[width=0.49\textwidth]{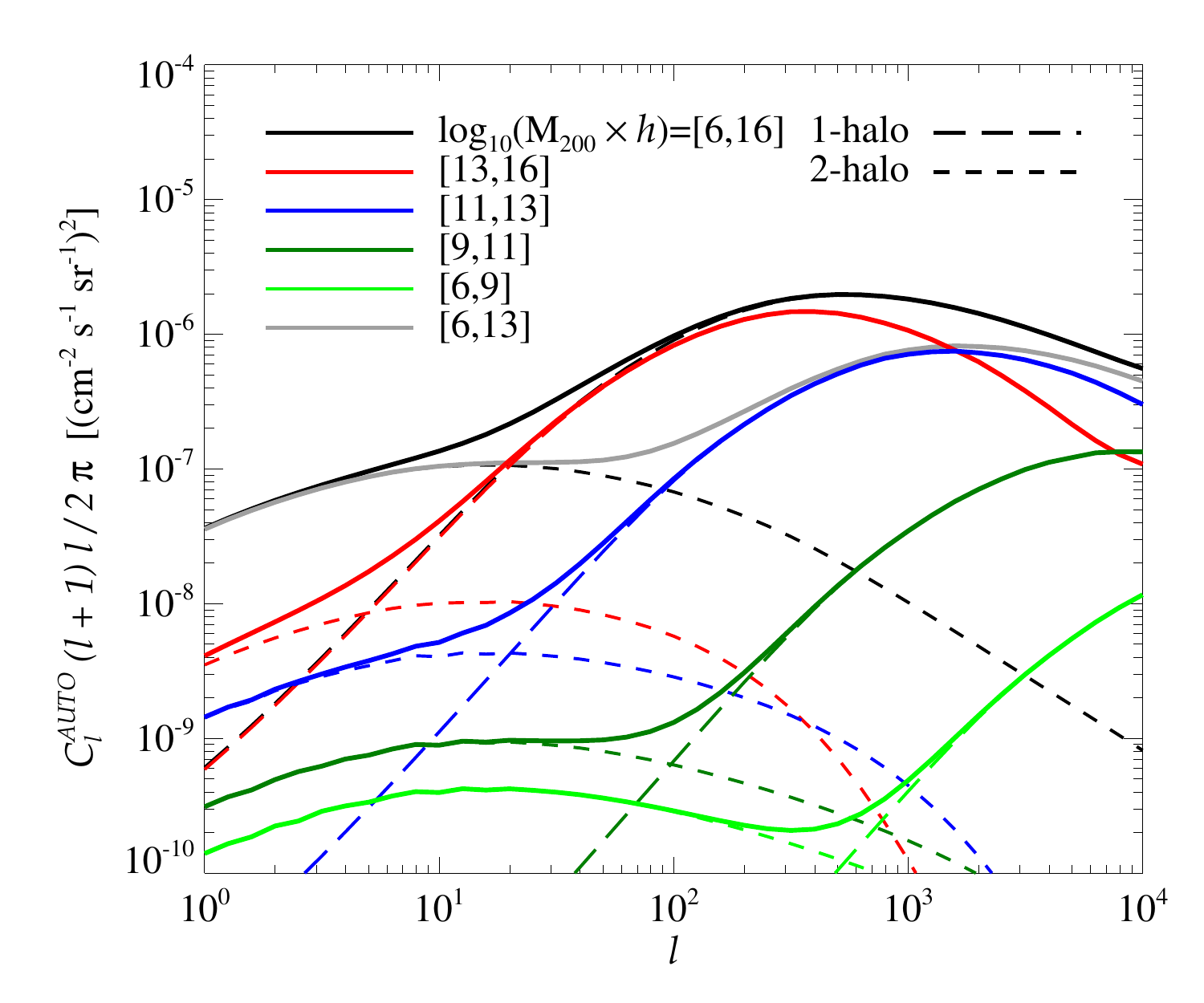}
  \caption{Contributions to the auto-correlation angular power spectrum for our
  reference sterile neutrino scenario split into different halo mass ranges. Note that for
  the mass range $10^{6}-10^{13}$~M$_\odot$, only the total (1-halo + 2-halo)
  is shown. Note that our $P_{\mathrm{lin}}$ is obtained for a CDM scenario, therefore
  the contribution below $10^9$~M$_\odot$ is likely overestimated \cite{2015arXiv150701998B}.
  However, as clear from the figure, this regime has no impact on our conclusions regarding
  the auto- and cross-correlation power spectra.}
  \label{fig:masses_contrib}
\end{figure}

We also try an additional approach to further reduce the contribution from
clusters.  This is simply to exclude completely clusters of galaxies from such
an analysis. One would think that by doing so, we would exclude most of the
sterile neutrino signal too, but we will see that the contrary is true. In
Fig.~\ref{fig:Cl_eRASSlimits}, we show how the contribution from clusters of
galaxies can be dramatically lowered when excluding clusters that will be
resolved by \emph{e}ROSITA. We consider both point-source and extended expected
sensitivities \cite{Merloni:2012} to bracket the case of extended and smaller,
far-away point-like clusters, from the first six months of survey (eRASS1) and
the complete survey (eRASS8). The left panel of Fig.~\ref{fig:Cl_eRASSlimits}
shows the auto-correlation power spectra, while the right panel shows the
cross-correlation with 2MRS. We compare these with the sterile neutrino
component, and with the same sterile neutrino component but integrated only up
to $M_{200}^{\nu, \mathrm{lim}} \times h = 10^{13}$~M$_\odot$, effectively excluding all clusters of
galaxies. The figures clearly show that excluding clusters of galaxies from
such an analysis does not have a big impact on the magnitude of the sterile
neutrino component, and, at the same time, that with the \emph{e}ROSITA survey
it will be realistic to exclude them as all will be resolved. The relative importance
of galaxy clusters becomes even clearer when looking at the individual contributions from 
different halo sizes in Fig.~\ref{fig:masses_contrib} where we show that a large
contribution to the sterile neutrino power spectra comes from galaxy-size halos. 
For completeness, we show the sterile neutrino component integrated only up to 
$M_{200}^{\nu, \mathrm{lim}} \times h = 10^{13}$~M$_\odot$ also in
the cross-correlation spectra of the above Fig.~\ref{fig:Cl_cross_AGN_gal}.

\begin{figure}[t]
  \centering
  \includegraphics[width=0.49\textwidth]{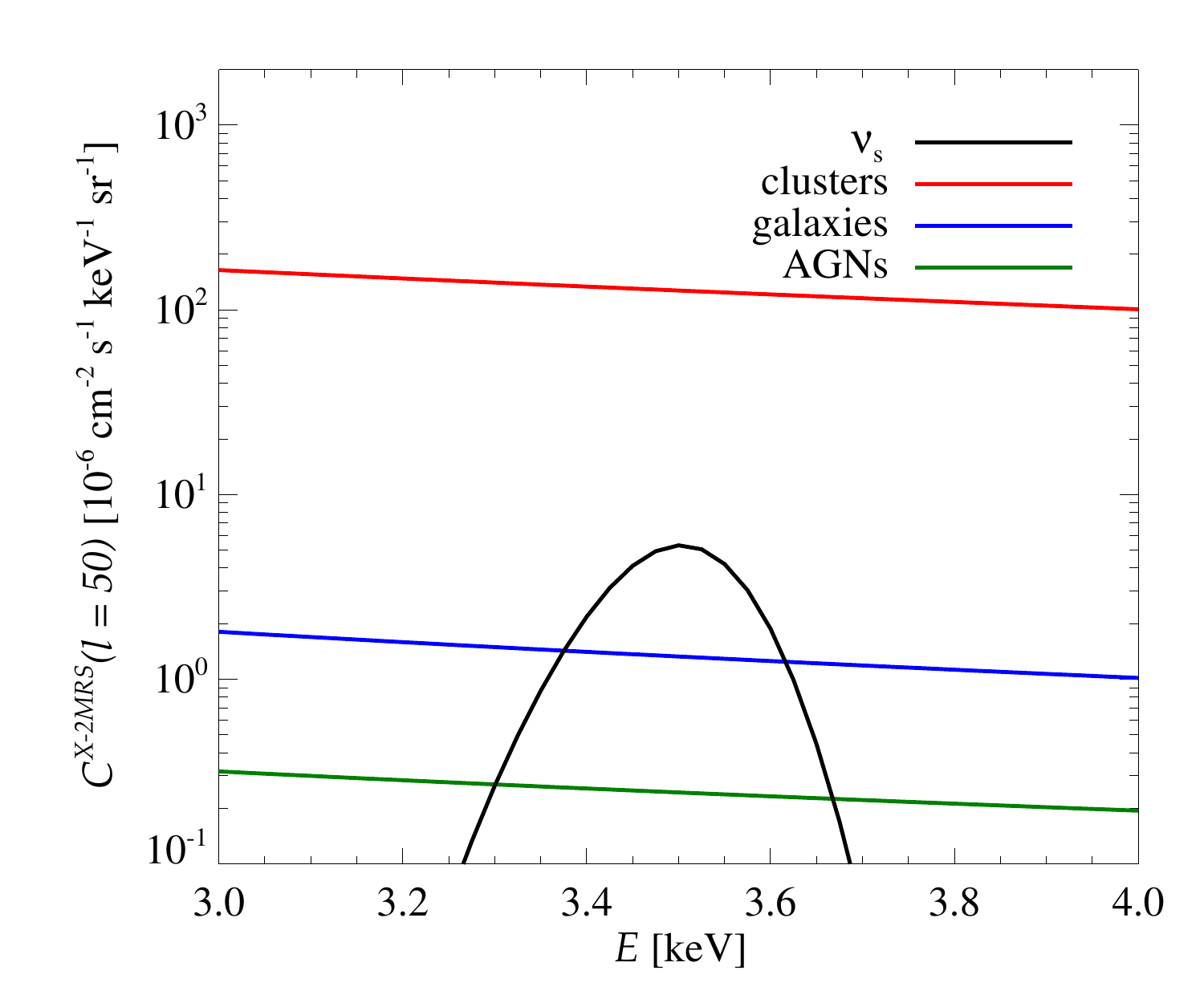}
  \includegraphics[width=0.49\textwidth]{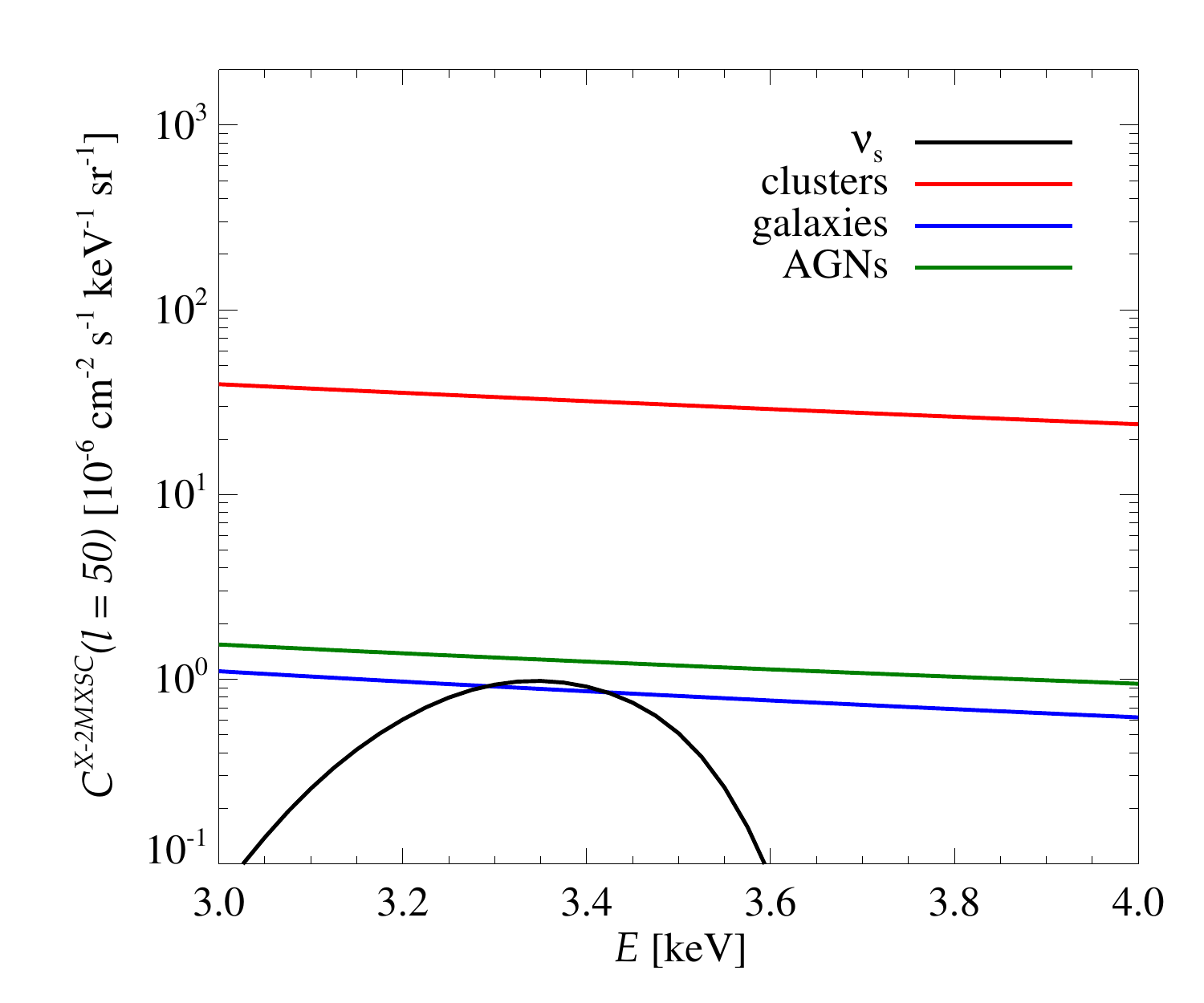}
  \caption{Angular cross-power spectrum for 2MRS (left) and 2MXSC (right) at
  $\ell = 50$ as a function of energy.}
  \label{fig:aps_energy}
\end{figure}

To conclude this section, in Fig.~\ref{fig:aps_energy}, we show the 2MRS (left)
and 2MXSC (right) cross correlation $C_\ell^{\times}(E)$ at $\ell = 50$ as a
function of energy. The cluster component dominates at all energies. The
sterile neutrino component dominates over those of AGNs and galaxies around its
peak energy for 2MRS, while it is just below for 2MXSC.  This is due not only
to the fact that high-redshift AGNs and galaxies cross-correlate more with the
higher redshifts of 2MXSC, as discussed for Fig.~\ref{fig:Cl_cross_AGN_gal},
but also to the fact that, in this case, the sterile neutrino decay line is
broader due to these higher redshifts too.  Looking back at the 2MXSC cross
correlation in the right panel of Fig.~\ref{fig:Cl_cross_AGN_gal}, we can argue
that a energy range broader than 3.4--3.6~keV, better including the sterile
neutrino decay line peak, would increase its cross-correlation with 2MXSC, but
would also increase that of the other components. Note that the choice of $\ell = 50$ 
for Fig.~\ref{fig:aps_energy} is arbitrary and done only with illustrative purposes. 
However, the behaviour at different multipoles is understandable noting the relative 
differences among different components in the cross-correlations of Fig.~\ref{fig:Cl_cross_AGN_gal}.
This is particularly true for the case of the cross-correlation with 2MRS where the
curves for the different components are remarkably similar at all scales.

Our main finding is that, although the sterile neutrino auto-correlation
component is completely dominated by clusters, AGNs and galaxies (see
Fig.~\ref{fig:Cl_auto}), by taking the cross correlation, we are able to
highlight it over the AGNs and galaxies. The contribution from clusters of
galaxies is very strong also in the cross-correlation case.  However, by
completely excluding clusters of galaxies from such an analysis, which is a
realistic assumption as \emph{e}ROSITA is expected to resolve them all, and
reducing the sterile neutrino component analysis only to structures up to
$M_{200}^{\nu, \mathrm{lim}} \times h = 10^{13}$~M$_\odot$, we do not loose a significant fraction of sterile
neutrino decays. In the next section, we will show that a detailed analysis
will be able to identify the sterile neutrino dark matter over the backgrounds
in both cases.

\section{Detectability and projected limits}
\label{sec:results}
In this section, we finally study the sensitivity for sterile neutrino signals
using the cross-correlation angular power spectrum. To this end, we perform a
$\chi^2$ fit to mock data in different energy ranges, with free normalisation
of the individual background components and of the signal. We take properly
into account the covariance of the different components, which is usually neglected.

As already mentioned, we take the soon-to-be-launched \emph{e}ROSITA as
reference X-ray satellite.  Therefore, we adopt an angular resolution $\sigma_b
= 15''$ and an (FWHM) energy resolution of $138\rm\,eV$.  The angular and
energy resolution have a mild dependence on energy in the range of interest
(see \cite{Merloni:2012} and Footnote~\ref{note1}), which we neglect for this
work.  The effective area is of the order of 500 cm$^2$ and the field of view
(FoV) around $55'$ diameter, with a strong dependence on energy, and we adopt
here the energy-dependent values for the grasp as given in~\cite{Merloni:2012}.
The total observation time is assumed to be four years.

\subsection{Statistical method}
The relevant uncertainties that we have to consider in our statistical analysis
come from the shot noise of the measured X-ray photons, the finite number of
measured galaxies, and the cosmic variance.  The X-ray shot noise affects
different X-ray energy ranges differently, whereas galaxy shot noise and cosmic
variance affect all energy ranges in the same way.

We perform a sideband analysis fitting the data in three energy ranges
simultaneously. In this case the cross-correlation angular power spectrum
$C_\ell^{\gamma, g}$ becomes a vector with three entries, and its variance is
described by a $3\times 3$ covariance matrix.  This matrix is given by (see
Appendix~\ref{apx} for details)
\begin{eqnarray}
  (\Delta C_{\ell}^{\gamma,{\rm g}})^2_{ij} &=& \frac{1}{(2\ell +1) f_{\rm sky}}
  \left[
    C_{\ell i}^{\gamma,g}
    C_{\ell j}^{\gamma,g}+
    \left(\frac{C_{Ni}^\gamma \delta_{ij}}{W_\ell^2} + 
      \sqrt{C_{\ell i}^\gamma C_{\ell j}^\gamma} 
    \right)
  \left(C_N^g + C_\ell^g \right)
  \right]
  \;,
  \label{eq:error2}
\end{eqnarray}
where $C_N^\gamma=4\pi\langle I \rangle^2 / N_\gamma$ and $C_N^g=4\pi/N_g$ are
respectively the photon and galaxy count shot noise terms, and $\langle
I\rangle$ is the average intensity of the total flux, as shown in
Fig.~\ref{fig:intensity}, in a given energy range.  The indices $i$ and $j$
refer to different energy ranges, and $\delta_{ij}$ is the Kronecker delta.
Furthermore, $f_{\rm sky}$ denotes the fraction of the sky that was observed,
and $W_\ell = \exp(-\sigma_b^2 \ell^2/2)$ is the detector window function of a
Gaussian point-spread function. Therefore, note that the contribution from the
photon shot-noise is diagonal, since the noise is independent for the various
energy ranges, and different in each because the mean intensity is different.
On the other hand, the galaxy shot noise affects the cross-correlation with
photons from different energy ranges in the same way.

We perform our analysis on three reference energies for the sterile neutrino
mass: $m_s =2.0$, 7.2 and 18.0~keV.  For each of the sterile neutrino masses,
we define three energy bands that are centred on the line, and are at slightly
lower and higher energies. These energy ranges are in keV units $\rm (low,
central, high)$ = ($0.5$--$0.8$, $0.9$--$1.1$, $1.2$--$1.5$), ($3.0$--$3.3$,
$3.4$--$3.6$, $3.7$--$4.0$) and ($8.5$--$8.8$, $8.9$--$9.1$, $9.2$--$9.5$).
We show the auto- and 2MRS cross-correlation spectra for the 
$0.9$--$1.1$ and $8.9$--$9.1$~keV energy bands in Appendix~\ref{apx2}.
The central band covers most of the signal as shown in the left panel of
Fig.~\ref{fig:aps_energy} for 2MRS. In the following, we will discuss results
only for the 2MRS catalogue for simplicity as the results that one would obtain
for the 2MXSC catalogue would be less constraining due to the different redshift
distribution of the latter as explained and showed in the previous sections.

We treat the instrumental backgrounds in a simple way, by increasing the photon
shot-noise term, $C^\gamma_{Ni}$, appropriately.  We adopt here the factors
5.0, 3.0 and 100 for 2.0, 7.2 and 18.0 keV sterile neutrinos, respectively.  At
these energies, the background is dominated by Galactic thermal emission, the
CXB, and the particle background, respectively~\cite{Merloni:2012}.

The $\chi^2$ function that we use for the sensitivity estimates is given by
\begin{equation}
  \chi^2 = \sum_{\ell=\ell_0}^{\ell_1} \sum_{i,j=1}^3
  (\bar C_{\ell i}^{\gamma,g} - C_{\ell i}^{\gamma,g}(\theta))
  \left[(\Delta C_{\ell}^{\gamma,g})^2\right]^{-1}_{ij}
  (\bar C_{\ell j}^{\gamma,g} - C_{\ell j}^{\gamma,g}(\theta))\;,
\end{equation}
where the first term in the first factor denotes the measured cross-correlation
spectrum, and the second term is the model.  The sum is taken over the angular
range $\ell = [\ell_0, \ell_1]$ and over all three energy bands.  Our default
angular range, to which we will refer as baseline case, is given 
by $\ell_0 = 10^2$ and $\ell_1=10^4$.

The model is a linear combinations of the contributions from sterile neutrinos,
unresolved AGNs and galaxies, and (resolved and unresolved) cluster emission.
Here, we are interested in deriving projected upper limits on a sterile
neutrino component.  We hence generate mock data (Asimov data) with the sterile
neutrino flux set to zero, and adopt the standard $\Delta\chi^2$ method to
derive $95\%$\,CL upper limits.  To this end, we increase the signal flux while
refitting the other parameters until the $\chi^2$ function changes by
$\Delta\chi^2=2.71$.

\subsection{Projected limits}
\begin{figure}
  \begin{center}
    \includegraphics[width=0.8\linewidth]{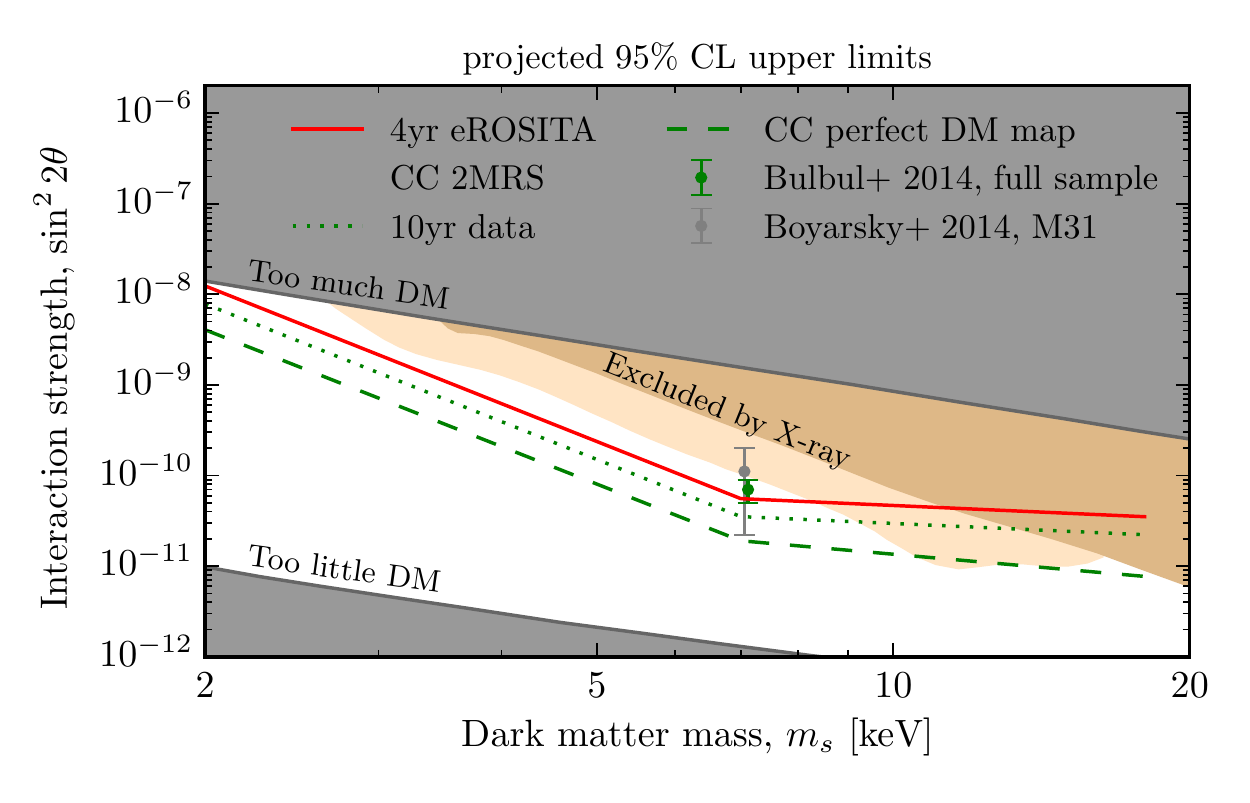}
  \end{center}
  \caption{Projected $95\%$ CL upper limits on the sterile neutrino mixing
  angle as function of the sterile neutrino mass.  For the baseline results
  (red solid line) we assume 4~yr of data taking with \emph{e}ROSITA, and a
  sideband analysis of the cross-correlation (CC) angular power spectrum with
  the 2MRS catalogue as discussed in the text. This is just on top of the
  tentative 3.5-keV line reported by different groups. For comparison, we also
  show the limits obtained for 10~yr of data taking (green dotted line), and
  when cross-correlating 4~yr data with a hypothetical \emph{perfect} model of
  the dark matter decay signal (dashed green line; no shot noise, same window
  function as for 2MRS and negligible bias). We derived projected limits for
  three reference energies and interpolate otherwise. For comparison, we also
  show regions that are excluded by over- or underproducing the observed DM relic
  density (grey areas), and regions excluded by previous X-ray observations
  (light brown regions), following Refs.~\cite{2012JCAP...03..018W,PhysRevD.89.025017,Boyarsky:2014jta}.}
  \label{fig:limits1}
\end{figure}

\begin{figure}
  \begin{center}
    \includegraphics[width=0.8\linewidth]{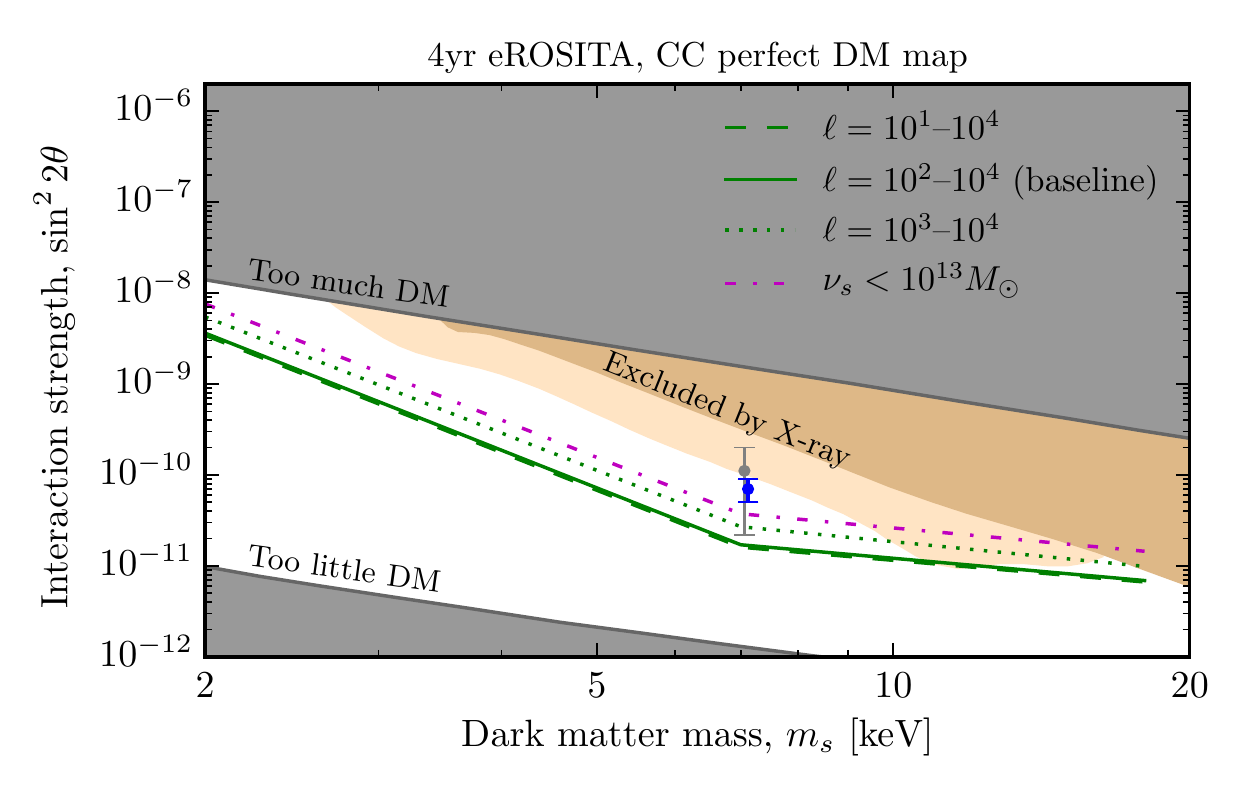}
  \end{center}
  \caption{Same as Fig.~\ref{fig:limits1}, for 4~yr of eROSITA data and
  cross-correlation (CC) with a perfectly modelled dark matter signal.  We show
  the impact of varying the considered angular range (green lines), finding
  that the impact of angular scales $\ell < 100$ is rather minor.  We also show
  how the limits change when masking out \emph{all} halos with masses above
  $M_{200} \times h = 10^{13}$~M$_\odot$ (magenta dash-dotted line).  
  The projected limits weaken then by a factor of two, but are then unaffected by the
  strong astrophysical line emission that is associated with the thermal
  emission from galaxy clusters. }
  \label{fig:limits2}
\end{figure}

The resulting projected limits are shown in Fig.~\ref{fig:limits1}. Most
interestingly, we find that for our baseline analysis \emph{e}ROSITA is
sensitive to our reference sterile neutrino scenario, which is motivated by the
findings of Refs.~\cite{Bulbul:2014sua,Boyarsky:2014jta}. If the claimed 3.5
keV feature is indeed due to dark matter decay, \emph{e}ROSITA should be able
to see it. 

We discuss now briefly what are the limiting factors for the sensitivity of our
baseline analysis, and how this could be further improved. In
Fig.~\ref{fig:limits1}, we also show the limits that would be obtained if
either the galaxy shot noise or the photon shot noise terms are set to zero.
These correspond, respectively, to the cases of either perfect knowledge about
the dark matter distribution (no shot noise, same window function as for 2MRS,
and negligible bias; note that the latter is close to one for galaxies, see
section~\ref{sec:autocorr}), or 10~yr observation time.  We find that the
galaxy noise term is the most limiting factor for current searches, suggesting
that improved measurements of the dark matter distribution in the local
Universe would be of great help when performing cross-correlation searches with
\emph{e}ROSITA data.

Finally, for comparison, we show in Fig.~\ref{fig:limits2} the limits that can
be obtained when masking out all massive clusters with masses above
$M_{200} \times h = 10^{13}$~M$_\odot$, for the cross-correlation with 
the hypothetical perfect DM tracer. The limits obtained in this way are 
only a factor $\lesssim2$ less constraining than what we found with our baseline analysis
(with the hypothetical perfect DM tracer). This suggest that a good sensitivity 
to dark matter signals can be obtained even if galaxy clusters, with potentially
problematic background lines from atomic transitions, are excluded from such an
analysis as we also showed in Figs.~\ref{fig:Cl_auto}, \ref{fig:Cl_cross_AGN_gal} 
and \ref{fig:masses_contrib}. Note, however, that in the current situation
where our knowledge of the local dark matter distribution is limited, the 
exclusion of the most massive halos from this analysis is not advisable as 
can be argued looking at our baseline model in Fig.~\ref{fig:limits1}: a factor
of 2 worsening could compromise the ability of testing the 3.5-keV line interpretation. 
We finally show that including larger angles (smaller $\ell$) in the analysis would not improve the
sensitivity further, whereas the ranges $\ell = 10^2$--$10^3$ and
$\ell=10^3$--$10^4$ are approximately equally important.

In general, we find that the cross-correlation analysis can significantly improve the
present constraints on sterile neutrinos below about $10$~keV, while above these energies 
the constraints that we obtain are comparable or worse than what can be obtained by pointed X-ray 
observations (e.g., from Refs.~\cite{2012JCAP...03..018W,PhysRevD.89.025017,Boyarsky:2014jta} as shown in 
Figs.~\ref{fig:limits1} and \ref{fig:limits2}) as our approach becomes dominated by the instrumental 
background (see also Fig.~\ref{fig:Cl_cross_AGN_gal_E} in Appendix~\ref{apx2}).

\section{Conclusions}
\label{sec:conc}
Motivated by claims of an unidentified 3.56-keV X-ray line found in several
galaxies and clusters of galaxies, and its interpretation in terms of the decay
of sterile neutrino dark matter~\cite{Bulbul:2014sua, Boyarsky:2014jta}, we
investigated for the first time the possibility of searching for such signatures using upcoming
full-sky observations of the cosmic X-ray background. We computed both the auto-
and cross-correlation angular power spectra, where for the latter we adopted a
nearby all-sky galaxy catalogue, for the sterile neutrino component and
astrophysical backgrounds.

As astrophysical components that give contributions to the X-ray background, we
considered AGNs, galaxies (powered by X-ray binaries), and thermal
bremsstrahlung emission from clusters of galaxies.  The combinations of these
three sources, in particular AGNs, give the dominant contribution to the mean
background intensity, and the component of the sterile neutrino decays, with
inferred parameters from \cite{Bulbul:2014sua, Boyarsky:2014jta}, gives only
$\sim$1\% contribution, even at the peak energy around 3.5~keV.

We found that the auto-correlation angular power spectrum is dominated by the
galaxy cluster component at almost all angular scales, as they are a rare (more
rare than AGNs and galaxies) and individually bright sources.  AGNs dominate
over galaxies and the sterile neutrino decays, and the latter gives only a
small contribution to the total angular power, comparable to that of galaxies.
Since $e$ROSITA will resolve most galaxy clusters through their thermal
emission, the cluster component could be in principle excluded from the
analysis.  Even in that case, however, a large contribution from the unresolved
AGNs would hide the underlying dark matter component almost completely.

Since the window function of the spectroscopic 2MASS galaxy catalogue has
considerable overlap with that of the sterile neutrino decays, taking the
cross-correlation with the 2MASS galaxies efficiently highlight the dark matter
component. In fact, we found that the cross-correlation power spectrum of
sterile neutrino decays is larger than that of both AGNs and galaxies. The
cluster component is still larger than the sterile neutrino one. However, when
masking all the clusters that will be resolved with $e$ROSITA, we found that
the dark matter component (computed from all the halos up to
$M_{200} \times h = 10^{13}$~M$_\odot$) becomes the largest 
in the cross-correlation power spectrum at almost all
angular scales.

\medskip

Using the above results, we performed a $\chi^2$ analysis in order to estimate
the expected sensitivity of {\it e}ROSITA for sterile neutrino searches.  We
perform a sideband analysis using three energy ranges --- one centred on the
potential line and other two, higher and lower, bands to calibrate continuous
astrophysical components. We find that when cross-correlating \emph{e}ROSITA
data with the 2MASS galaxies, the projected 95\%~CL upper limits after four
years of observation are cutting into the parameter region of the recently
claimed 3.56~keV sterile neutrino decay line~\cite{Bulbul:2014sua,
Boyarsky:2014jta}. Moreover, we show that the cross-correlation analysis 
can significantly improve the present constraints on sterile neutrinos below about $10$~keV, 
while above these energies our approach becomes dominated by the instrumental background
and the corresponding constraints are not as competitive with respect to pointed X-ray observations.
We also find that the masking of galaxy clusters does not
improve the constraints, unless a better knowledge of the local dark
matter distribution is achieved. In fact, the main limitation comes from the shot noise associated
with the finite number of galaxies in the 2MASS catalogue. Indeed, we showed
that when using shot-noise free tracers for dark matter, i.e., having a perfect
knowledge of the local dark matter distribution with no bias, the projected
limits improve significantly, allowing \emph{e}ROSITA to put stringent
constraints on the possible sterile neutrino signal. \emph{Hence, in order to
ensure that the potential of \emph{e}ROSITA and other full-sky X-ray surveys is
fully realised for dark matter searches, it is critical to obtain a detailed
description of the dark matter distribution at low redshift, at galaxy and
galaxy cluster scales, without the shot-noise limitations of the 2MASS
catalogue.} To what extent this can be achieved by a combination of optical and
X-ray observations (e.g.,~Ref.~\cite{2014FrP.....2....6F}), and of gravitational
lensing~\cite{2013ApJ...771L...5C, Shirasaki:2014noa} will be discussed in
a forthcoming publication~\cite{forthcoming}.

\bigskip

\paragraph{Note added.}
In the final stages of our work we became aware of another group exploring a
similar approach~\cite{CampbellPrep}.

\acknowledgments
We would like to thank Andrea Chiappo, Marco Regis, Miguel-Angel Sanchez-Conde 
and Petr Tinyakov for useful discussions. FZ acknowledge the support of the Spanish 
MICINNs Consolider-Ingenio 2010 Programme under grant MultiDark CSD2009-00064, 
AYA10-21231. This work was supported by the Netherlands Organization for Scientific 
Research (NWO) through one Veni (FZ) and two Vidi (SA and CW) grants.

\appendix
\section{Treatment of correlated statistical uncertainties}
\label{apx}
We discuss briefly how the covariance matrix used in our $\chi^2$ analysis
follows from first principles.  For a discussion about the angular power
spectrum in light of finite counts see also~\cite{Campbell:2014mpa}. 

Our analysis is based on the cross correlation angular power spectrum that is
obtained from the measurements in some X-ray energy band $i$ and the galaxy
catalogue $g$ (mostly 2MRS), and given by
\begin{equation}
  C_{\ell i}^{\gamma, g} \equiv \langle a^{\gamma}_{\ell m i} (a_{\ell
  m}^g)^\ast \rangle \equiv \frac{1}{2\ell+1} \sum_{m=-\ell}^\ell
  a^{\gamma}_{\ell m i} (a_{\ell m}^g)^\ast\;.
\end{equation}
Here, $a_{\ell m i}^\gamma$ ($a_{\ell m}^g$) are the modes of a decomposition
of the \emph{measured} X-ray sky map (galaxy catalogue) into spherical harmonics,
and $\langle \dots\rangle$ denotes the usual average over $m$.

We are interested in the variations of $C_{\ell i}^{\gamma,g}$ over many
measurements.  In general, it holds that  $\langle a_{\ell m i}^\gamma\rangle_r
= \langle a_{\ell m}^g\rangle_r = 0$, where $\langle \dots \rangle_r$ is the
average over cosmological realisations, and photon and galaxy shot noise.
Furthermore, we find that
\begin{eqnarray}
  \langle a_{\ell m i}^\gamma (a_{\ell m j}^\gamma)^\ast \rangle_r 
  &=& \frac{4\pi}{N_{\gamma, i}}\langle I \rangle_i^2 \delta_{ij} + \sqrt{
  C_{\ell i}^\gamma C_{\ell j}^\gamma}\;,
  \\
  \langle a_{\ell m i}^\gamma (a_{\ell m }^g)^\ast \rangle_r
  &=& C_{\ell i}^{\gamma,g} \;,
  \\
  \langle a_{\ell m }^g (a_{\ell m }^g)^\ast \rangle_r 
  &=& \frac{4\pi}{N_{g}} + C_{\ell}^g\;,
\end{eqnarray}
where $N_{\gamma,i}$ is the number of photons in energy bin $i$, and $\langle I
\rangle_i$ the corresponding full-sky averaged flux.  The noise comes either
from the finite number of photons in X-ray observations or the finite number of
galaxies in the catalogue, and the cosmic variance contribution is a
consequence of the stochasticity of structure formation.  We assumed that the
morphology of the signal changes only weakly between the relevant energy bins.

The covariance of the cross correlation angular power spectrum is then obtained
as
\begin{eqnarray}
  (\Delta C_{\ell}^{\gamma, {\rm g}})^2_{ij}
  &\equiv& \langle C_{\ell i}^{\gamma, {\rm g}} C_{\ell j}^{\gamma, {\rm g}}
  \rangle_r \\\nonumber
  & = & \frac{1}{(2\ell+1)^2}  
  \sum_{m,m'=-\ell}^\ell
  \langle a_{\ell mi}^\gamma (a_{\ell m'}^g)^\ast
  a_{\ell m'j}^\gamma (a_{\ell m}^g)^\ast 
  \rangle_r
  - \langle a_{\ell mi}^\gamma (a_{\ell m}^g)^\ast \rangle_r
  \langle a_{\ell m'j}^\gamma (a_{\ell m'}^g)^\ast \rangle_r\\\nonumber
  & = & \frac{1}{(2\ell+1)^2}  
  \sum_{m=-\ell}^\ell
  \langle a_{\ell m i}^\gamma (a_{\ell m j}^\gamma)^\ast \rangle_r 
  \langle a_{\ell m }^g (a_{\ell m }^g)^\ast \rangle_r  +
  \langle a_{\ell m i}^\gamma (a_{\ell m }^g)^\ast \rangle_r
  \langle a_{\ell m j}^\gamma (a_{\ell m }^g)^\ast \rangle_r \\\nonumber
  & = & \frac{1}{(2\ell+1)} \left[
    \left(\frac{4\pi}{N_{\gamma, i}}\langle I \rangle_i^2 \delta_{ij} +
      \sqrt{ C_{\ell i}^\gamma C_{\ell j}^\gamma}\right)
    \left( \frac{4\pi}{N_{g}} + C_{\ell}^g \right) +
  C_{\ell i}^{\gamma,g} C_{\ell j}^{\gamma,g} \right]\;.
\end{eqnarray}
The adopted Eq.~\eqref{eq:error2} is then obtained after a rescaling of the
photon power spectrum that inverts the (here energy-independent) effect of the
finite angular resolution, e.g.,~$C^\gamma_{\ell i}\to C^\gamma_{\ell
i}/W_\ell^2$.

\section{Auto- and cross-correlation power spectra}
\label{apx2}
\begin{figure}
  \begin{center}
    \includegraphics[width=0.49\textwidth]{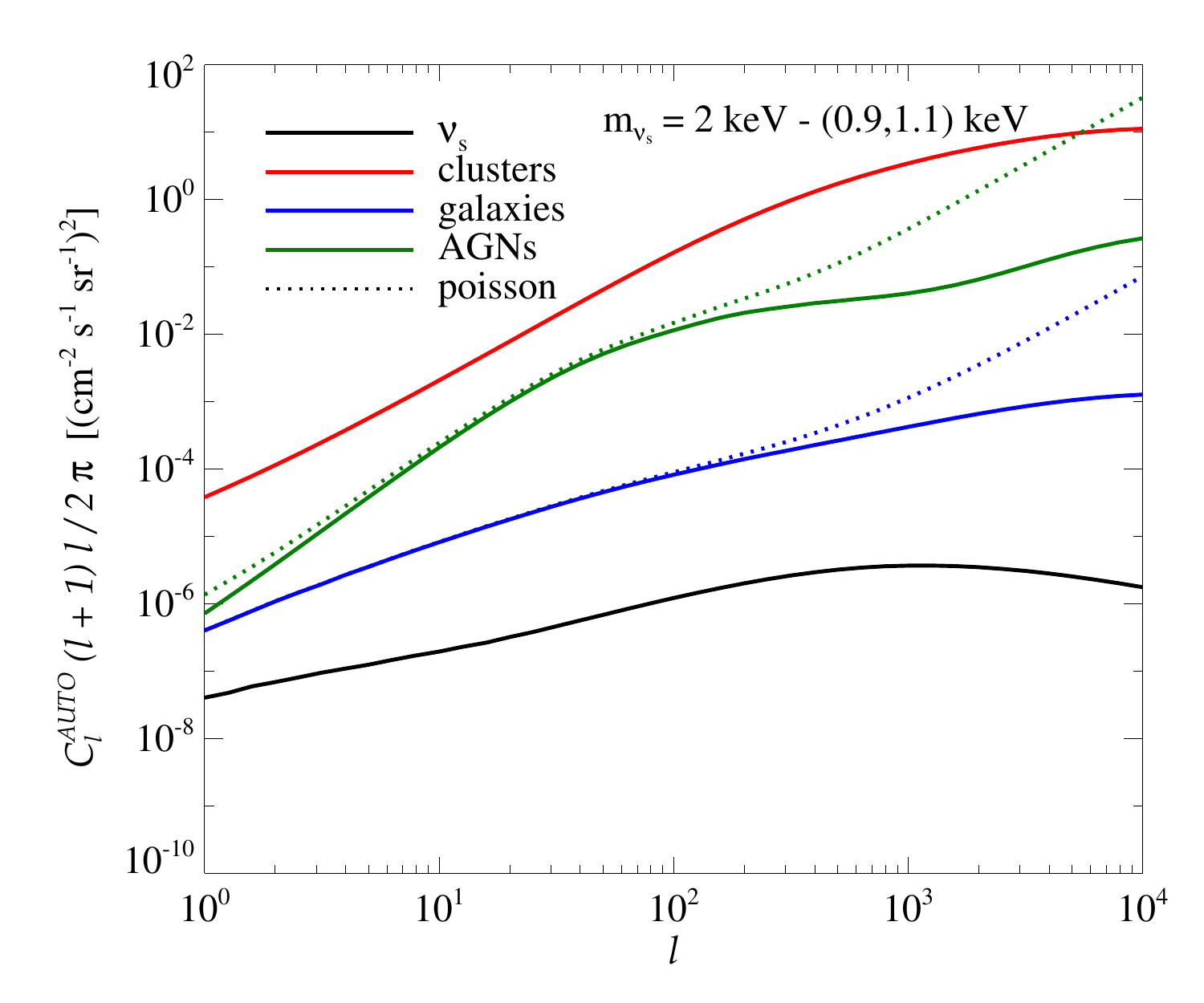}
    \includegraphics[width=0.49\textwidth]{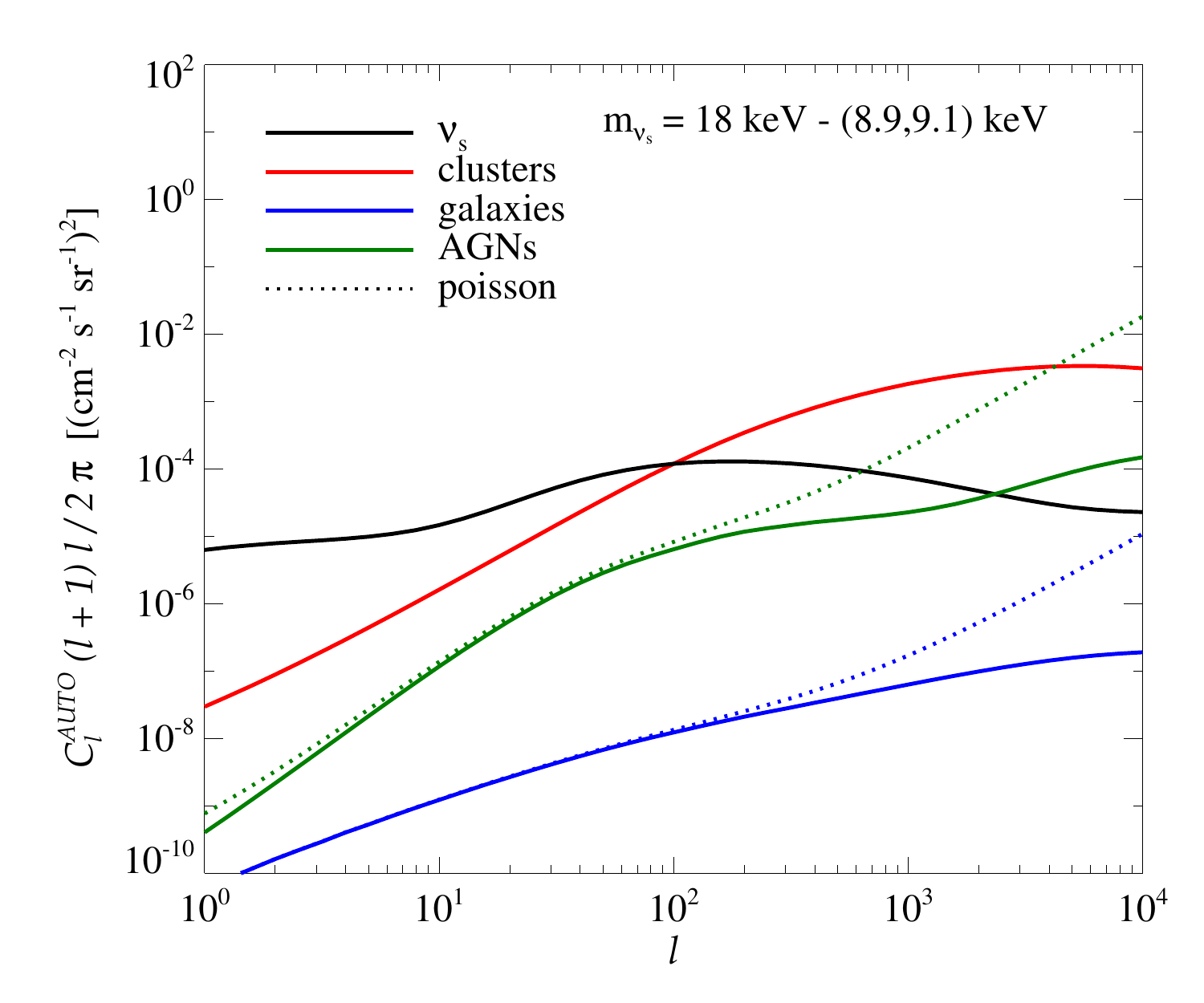}
    \caption{\textbf{Left.} Auto-correlation angular power spectrum of CXB due
    to sterile neutrino decays for $m_{\nu_s} = 2$~keV with a mixing angle
    corresponding to the constraint given by our baseline analysis as in Fig.~\ref{fig:limits1}, 
    (unresolved) AGNs, (unresolved) galaxies, and (resolved and unresolved) clusters of galaxies
    in the $0.9-1.1$~keV energy band. For AGNs and galaxies, both the
    correlation term (solid) and total including Poisson term (dotted) are
    shown. \textbf{Right.} The same but for $m_{\nu_s} = 18$~keV in the
    $8.9-9.1$~keV energy band. We show 
    both plots with the same scale for comparison, but note that the scale is
    different with respect to the left panel of Fig.~\ref{fig:Cl_auto}.}
    \label{fig:Cl_auto_E}
  \end{center}
\end{figure}

\begin{figure}
  \centering
  \includegraphics[width=0.49\textwidth]{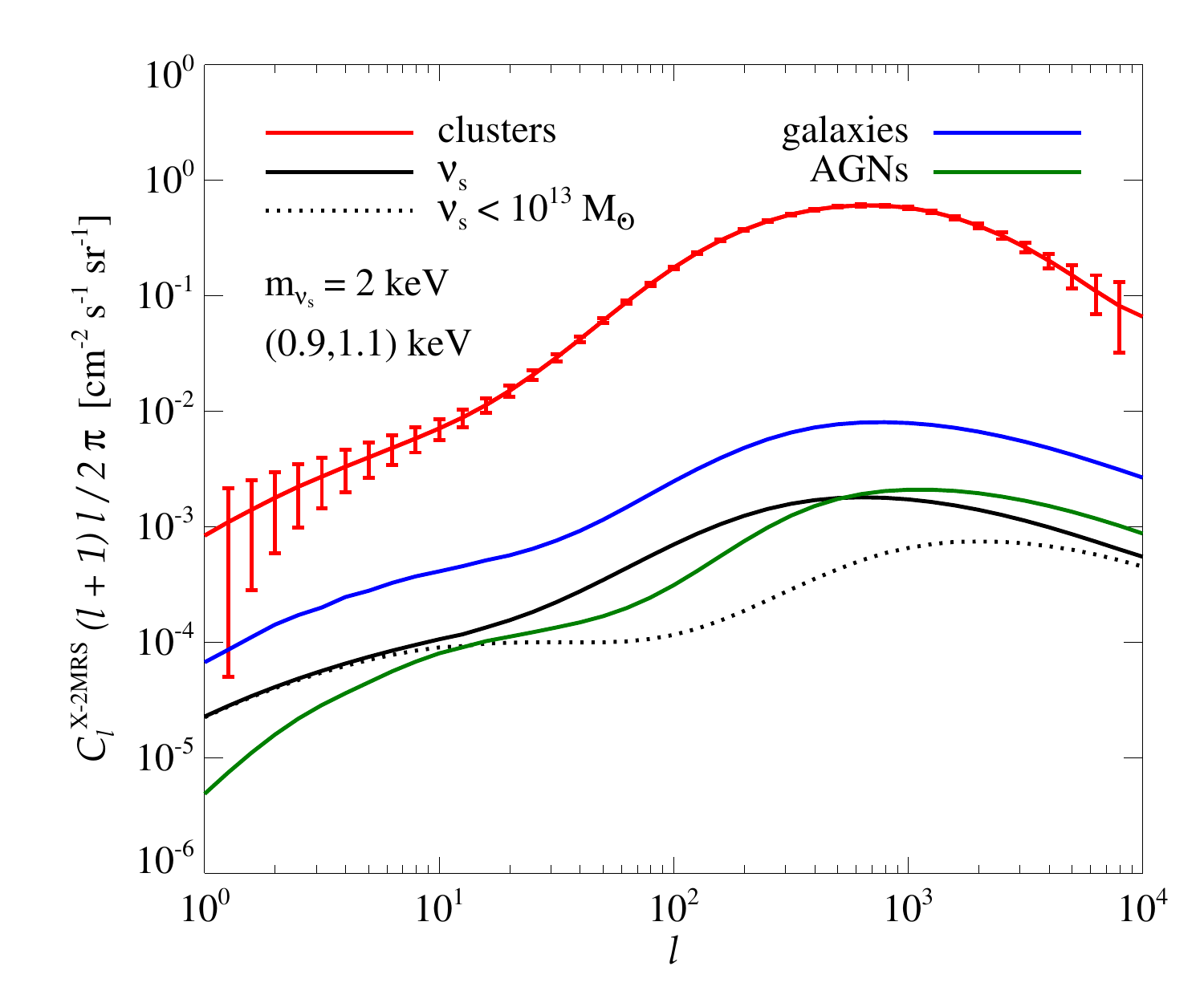}
  \includegraphics[width=0.49\textwidth]{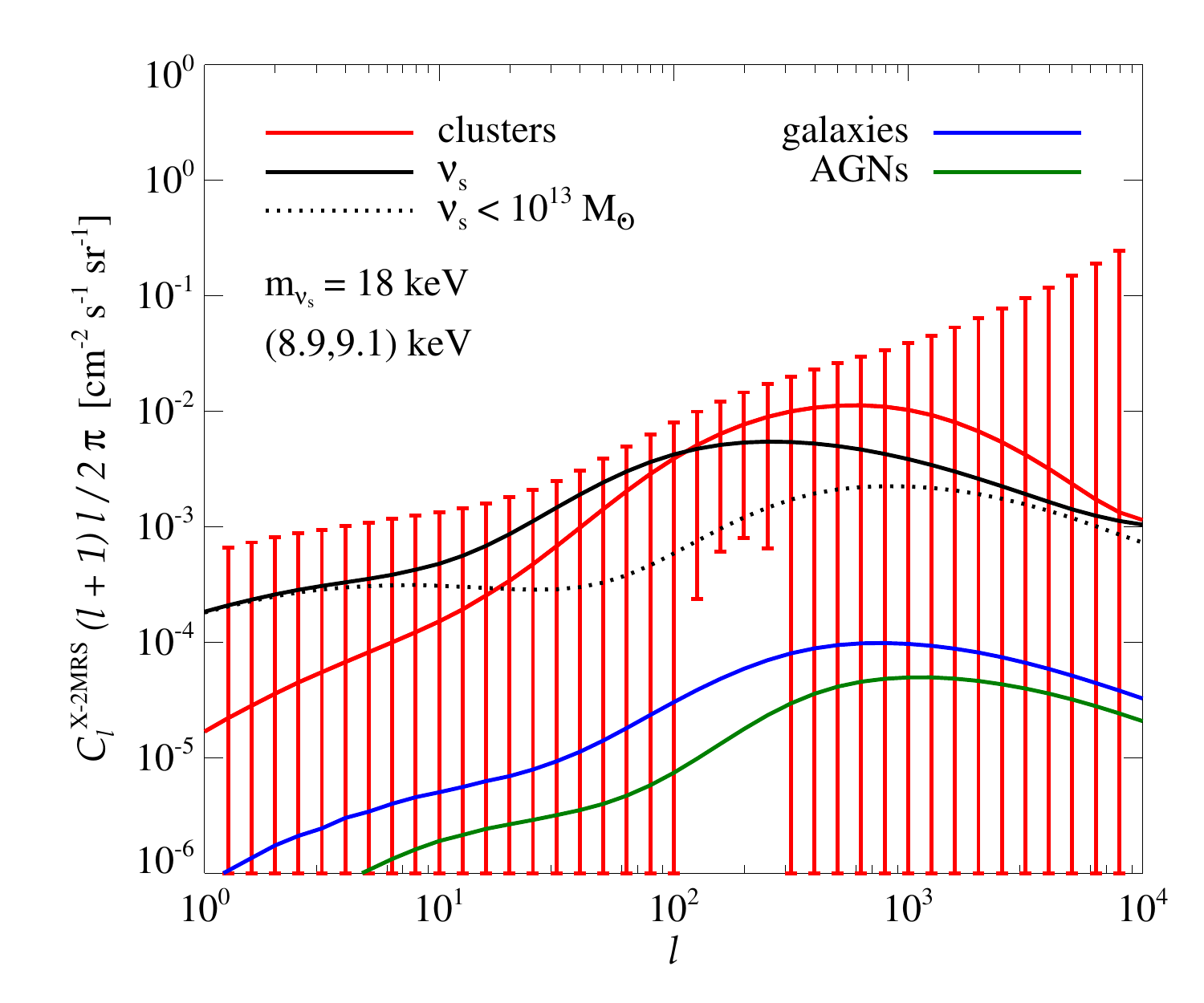}
  \caption{Cross correlation with 2MRS. We overlay to the cluster component the 
  error bars, as $\Delta\,C_{l}/\sqrt{l/2}$, estimated from the diagonal part of the covariance matrix of Eq.~\ref{eq:error2}. 
  {\bf Left.} Cross-correlation angular power spectrum 
  of CXB due to sterile neutrino decays  for $m_{\nu_s} = 2$~keV with a mixing angle
  corresponding to the constraint given by our baseline analysis as in Fig.~\ref{fig:limits1}, 
  (unresolved) AGNs, (unresolved) galaxies, and (resolved and unresolved) clusters of galaxies 
  in the $0.9-1.1$~keV energy band. For sterile neutrinos, we also show the case when the
  integration upper mass limit is fixed to $M_{200}^{\nu, \mathrm{lim}} \times h = 10^{13}$~M$_\odot$. See
  main text for details. {\bf Right.} The same but for $m_{\nu_s} = 18$~keV in the
  $8.9-9.1$~keV energy band. We show both plots with the same scale for comparison, but note that the scale is
  different with respect to the left panel of Fig.~\ref{fig:Cl_cross_AGN_gal}.}
  \label{fig:Cl_cross_AGN_gal_E}
\end{figure}

We show in Figs.~\ref{fig:Cl_auto_E} and \ref{fig:Cl_cross_AGN_gal_E} the
auto-correlation and 2MRS cross-correlation power spectra in the $0.9-1.1$ and 
$8.9-9.1$~keV energy ranges corresponding to the sterile neutrino masses of $m_{\nu_s} = 2$
and $18$~keV, respectively. In both cases, the sterile neutrino mixing angle is chosen to be just as
from the constraints of our baseline analysis shown in Fig.~\ref{fig:limits1}. Our constraining
power decrease dramatically moving toward higher energies, as clear both from Fig.~\ref{fig:limits1} and
from the error bars in the left panel of Fig.~\ref{fig:Cl_cross_AGN_gal_E}, due to the
larger instrumental background.

\bibliographystyle{JHEP}
\bibliography{Xline.bib}

\end{document}